\documentclass[10pt, conference]{IEEEtran}
\IEEEoverridecommandlockouts
\usepackage{cite}
\usepackage{amsmath,amssymb,amsfonts}
\usepackage{algorithmic}
\usepackage{graphicx}
\usepackage{textcomp}
\usepackage{xcolor}
\usepackage{color,soul}
\usepackage{subfig}
\def\BibTeX{{\rm B\kern-.05em{\sc i\kern-.025em b}\kern-.08em
    T\kern-.1667em\lower.7ex\hbox{E}\kern-.125emX}}
    
\makeatletter
\newcommand{\thickhline}{%
    \noalign {\ifnum 0=`}\fi \hrule height 1pt
    \futurelet \reserved@a \@xhline
}

\graphicspath{{image/}}

\usepackage{caption}
\usepackage{mathptmx} 
\usepackage{hyperref}
\usepackage{amsmath}
\usepackage{fancyhdr}
\usepackage{lipsum}
\usepackage{setspace}

\usepackage{color}
\usepackage{listings}
\usepackage{alphalph}

\usepackage{pbox}
\usepackage{makecell} %
\usepackage{fancyhdr}
\usepackage{lipsum}
\usepackage{pifont}
\newcommand{\cmark}{\ding{51}}%
\newcommand{\xmark}{\ding{55}}%

\newcommand{\projname}{UVMBench}

\begin{document}
\bstctlcite{IEEEexample:BSTcontrol}
\title{\projname: A Comprehensive Benchmark Suite for Researching Unified Virtual Memory in GPUs}

\author{\IEEEauthorblockN{Yongbin Gu, Wenxuan Wu, Yunfan Li and Lizhong Chen}
\IEEEauthorblockA{\textit{School of Electrical Engineering and Computer Science} \\
\textit{Oregon State University, Corvallis, USA}\\
\{guyo, wuwen, liyunf, chenliz\}@oregonstate.edu}
}

\maketitle

\begin{abstract}
	The recent introduction of Unified Virtual Memory (UVM) in GPUs offers a new programming model that allows GPUs and CPUs to share the same virtual memory space, which shifts the complex memory management from programmers to GPU driver/ hardware and enables kernel execution even when memory is oversubscribed. Meanwhile, UVM may also incur considerable performance overhead due to tracking and data migration along with special handling of page faults and page table walk. 
	As UVM is attracting significant attention from the research community to develop innovative solutions to these problems, in this paper, we propose a comprehensive UVM benchmark suite named \textit{UVMBench} to facilitate future research on this important topic. The proposed \textit{UVMBench} consists of 32 representative benchmarks from a wide range of application domains. The suite also features unified programming implementation and diverse memory access patterns across benchmarks, thus allowing thorough evaluation and comparison with current state-of-the-art. A set of experiments have been conducted on real GPUs to verify and analyze the benchmark suite behaviors under various scenarios. 
	
\end{abstract}


\section{Introduction}
\label{introduction}
GPUs have been gaining great attention in accelerating traditional and emerging workloads, such as machine learning, bioinformatics, electrodynamics, etc. due to GPU's massively parallel computing capability. However, there are two major issues in the mainstream GPU programming model that severely limit further utilization. First, the physical memory separation between a GPU and a CPU requires explicit memory management in conventional GPU programming model. Programmers have to explicitly copy data between CPU and GPU memories to the location where the data is used (i.e. \textit{copy-then-execute}). Second, the conventional GPU programming model does not allow a kernel to be executed if it needs more memory that what the GPU memory can provide (i.e., \textit{memory oversubscription}). This has greatly limited the use of GPUs in large data-intensive machine learning applications \cite{devlin2018bert,wu2019pointconv} nowadays. Recently, GPU vendors have proposed and started to employ a new approach, \textit{Unified Virtual Memory (UVM)}, in the newly released products\cite{sakharnykh,AMD}. UVM allows GPUs and CPUs to share the same virtual memory space, and offloads memory management to the GPU driver and hardware, thus eliminating explicit copy-then-execute by the programmers. The GPU driver and underlying hardware automatically migrate the needed data to destinations. Moreover, UVM enables GPU kernel execution while memory is oversubscribed by automatically evicting data that is no longer needed in the GPU memory to the CPU side. This is extremely important and helpful in facilitating large workloads (especially deep learning models) and GPU virtualization \cite{lu2019gqos, gu2019low} with limited memory sizes.

However, the advantages of UVM may come at a price. Analogous to virtual machines that offer great flexibility over physical machines but sacrifice performance in some degree \cite{zhang2012performance}, UVM also incurs performance overhead. In order to implement automatic data migration between a CPU and a GPU, the GPU driver and the GPU Memory Management Unit (MMU) have to track data access information and determine the granularity of data migration over the PCIe link \cite{ganguly2019interplay}. This may reduce performance. For example, UVM needs special page table walk and page fault handling that introduce extra latency for memory accesses in GPUs. In addition, the fluctuated page migration granularity may also under-utilize PCIe bandwidth. 



Due to the large potential benefits of UVM and its associated performance issues, UVM has recently drawn significant attention from the research community. Several optimization techniques have been proposed to mitigate the side effects of UVM \cite{zheng2016towards, markthub2018dragon, li2019framework, ganguly2019interplay, yu2019quantitative, kim2020batch, gangulyadaptive}. The earliest work is Zheng \textit{et al.} \cite{zheng2016towards}, which enables on-demand GPU memory and proposes prefetching techniques to improve UVM performance. As the work predates the release of UVM, the developed on-demand memory APIs are quite different from the version in the current UVM practice. More recently, Ganguly \textit{et al.} \cite{ganguly2019interplay}, Yu \textit{et al.} \cite{yu2019quantitative} and Li \textit{et al.} \cite{li2019framework} study prefetching and/or eviction techniques for UVM in more detail. However, their evaluation includes only benchmarks with limited number of access patterns, which makes it difficult to assess the effectiveness of their schemes on a broader range of benchmarks with diverse memory access patterns. 
In fact, comprehensive benchmarks (or the lack thereof) have become a common issue in these and other prior works on GPU UVM. 
Most of them have used their own modified versions of existing benchmark suites (e.g., Rodinia \cite{che2009rodinia, che2010characterization}, Parboil \cite{stratton2012parboil}, Polybench \cite{pouchet2012polybench}) or several in-house workloads. Our further inspection of these benchmarks shows that they lack unified implementation and no paper so far has provided a thorough analysis of the memory behaviors of these benchmarks. This can be a serious limitation for researchers and developers who aim to propose new optimizations for UVM and who would like to make comparison with existing research works. 

\begin{table*}[]
	\centering
	\small
	\caption{List of Benchmarks in the proposed UVMBench.}
	\begin{tabular}{|c|c|c|c|c|c|}
		\hline
		\textbf{Application}                               & \textbf{Abbr.}      & \textbf{Domain}             & \textbf{Kernels} & \textbf{Threads Per Block} & \textbf{Type} \\ \hline
		2D Convolution                            & 2DCONV     & Machine Learning   & 1       & 256               & R    \\
		2   Matrix Multiplications                & 2MM        & Linear Algebra     & 2       & 256               & R    \\
		3D   Convolution                          & 3DCONV     & Machine Learning   & 1       & 256               & R    \\
		3   Matrix Multiplications                & 3MM        & Linear Algebra     & 3       & 256               & R    \\
		Matrix Transpose  Vector   Multiplication & ATAX       & Linear Algebra     & 2       & 256               & I    \\
		Backpropgation                            & BACKPROP   & Machine Learning   & 2       & 256               & R    \\
		Breath First Search                       & BFS        & Graph Theory       & 6       & 1024              & I    \\
		BiCGStab Linear Solver                    & BICG       & Linear Algebra     & 2       & 256               & I    \\
		Bayesian Network                          & BN         & Machine Learning   & 2       & 256               & R    \\
		Convolution Neurak Network                & CNN        & Machine Learning   & 6       & 64                & R    \\
		Correlation Computation                   & CORR       & Statistics         & 4       & 256               & I    \\
		Covariance Computation                    & COVAR      & Statistics         & 3       & 256               & I    \\
		Discrete Wavelet Transform 2D             & DWT2D      & Media Compression  & 2       & 256               & R    \\
		2-D Finite Different Time Domain          & FDTD-2D    & Electrodynamics    & 3       & 256               & I    \\
		Gaussian Elimination                      & GAUSSIAN   & Linear Algebra     & 2       & 512/16            & I    \\
		Matrix-multiply                           & GEMM       & Machine Learning   & 1       & 256               & I    \\
		Scalar, Vector Matrix Multiplication      & GESUMMV    & Machine Learning   & 1       & 256               & I    \\
		Gram-Schmidt decomposition                & GRAMSCHM   & Linear Algebra     & 3       & 256               & I    \\
		HotSpot                                   & HOTSPOT    & Physics Simulation & 1       & 256               & R    \\
		HotSpot 3D                                & HOTSPOT3D  & Physics Simulation & 1       & 256               & R    \\
		Kmeans                                    & KMEANS     & Machine Learning   & 5       & 1/3               & I    \\
		K-Nearest Neighbors                       & KNN        & Machine Learning   & 4       & 256               & R    \\
		Logistic Regression                       & LR         & Machine Learning   & 1       & 128               & R    \\
		Matrix Vector Product    Transpose        & MVT        & Linear Algebra     & 2       & 256               & I    \\
		Needleman-Wunsch                          & NW         & Bioinformatics     & 2       & 16                & I    \\
		Particle Filter                           & PFILTER    & Medical Imaging    & 1       & 128               & R    \\
		Pathfinder                                & PATHFINDER & Grid Traversal     & 1       & 256               & R    \\
		Speckle Reducing Anisotropic Diffusion    & SRAD       & Image Processing   & 2       & 256               & R    \\
		Stream Cluster                            & SC         & Data Mining        & 1       & 512               & I    \\
		Support Vector Machine                    & SVM        & Machine Learning   & 2       & 1024              & I    \\
		Symmetric rank-2k operations              & SYR2K      & Linear Algebra     & 1       & 256               & I    \\
		Symmetric rank-k operations               & SYRK       & Linear Algebra     & 1       & 256               & R    \\ \hline
	\end{tabular}
	\label{benchname}
\end{table*}

In this paper, we aim to enrich the GPU UVM research community by developing a comprehensive UVM benchmark suite consisting of 32 representative benchmarks belonging to different application domains. This suite features unified programming implementation and diverse memory access patterns across benchmarks, allowing researchers to thoroughly evaluate and compare with current state-of-the-art. In addition to traditional benchmarks, the proposed suite also includes more machine learning related workloads, as GPUs have been increasingly used in machine learning tasks. This would help researchers to understand better the role that GPU UVM plays in machine learning acceleration.

The developed benchmarks are evaluated on a Nvidia GTX 1080 Ti GPU with 11GB memory capacity. The code volume is reduced by removing explicit memory management APIs thanks to UVM. Evaluation results show that, if we directly implement/convert benchmarks to the UVM programming model, there is an average of 34.2\% slowdown than the non-UVM benchmarks. However, if we augment with proper \textit{manual} optimizations on data prefetching and data reuse, the performance can be restored to almost the same as the non-UVM programming model. This indicates that there is substantial room for UVM research on developing \textit{autonomous} memory management to close the gap between UVM and non-UVM models and possibly exceed the performance of non-UVM. 
Our experiment also verifies the capability of the UVM-enabled benchmarks to execute successfully under memory oversubscription scenarios, where UVM essentially creates the illusion of a large GPU memory by using a small GPU memory and the CPU memory. While performance degradation is observed compared with a true large GPU memory, this enabling technology opens up new opportunities in accelerating large workloads on GPUs.

The main contributions of this paper are the following:

\begin{itemize}
	\item Identifying the need for a benchmark suite for UVM;
	\item Developing a comprehensive UVM benchmark suite to facilitate the research on UVM;
	\item Profiling memory access patterns of the benchmark suite, and studying the relevance of the patterns to performance under memory oversubscription;
	\item Conducting thorough analysis of performance difference between the UVM and non-UVM programming models.
\end{itemize}

\begin{table*}[]
	\centering
	\small
	\caption{UVMBench vs. other benchmarks or benchmark suites.}
	\begin{tabular}{|c|c|c|c|c|c|}
		\hline
		\makecell{\textbf{Benchmarks / Benchmark Suite}} & \makecell{\textbf{\# of Workloads}} & \makecell{\textbf{Test in Real} \\ \textbf{Hardware}} & \makecell{\textbf{Machine Learning} \\ \textbf{Workloads}} & \makecell{\textbf{Diverse Memory} \\ \textbf{Access Patterns}} & \makecell{\textbf{Oversubscription} \\ \textbf{Support}} \\ \hline
		Workloads in \cite{ganguly2019interplay} & 14 & \xmark & \xmark & \xmark & \cmark \\ \hline
		Workloads in \cite{chien2019performance}           & 6  & \cmark & \xmark & \xmark & \cmark \\ \hline
		Nvidia SDK \cite{NVSDK}  & 1 & \cmark & \xmark & \xmark & \xmark \\ \hline
		\textbf{UVMBench}  & \textbf{32} &\textbf{\cmark}  & \textbf{\cmark} & \textbf{\cmark} &\textbf{ \cmark} \\ \hline
	\end{tabular}
	\label{bench_comp}
\end{table*}

We have discussed the importance of GPU UVM research and the motivation for a benchmark suite in this section. In the remaining of this paper,
Section \ref{benchmark} describes the proposed benchmark suite in more detail. Section \ref{methodology} explains our evaluation methodology. Section \ref{results} presents and analyzes test results. Key observations drawn from the results and suggestions for future UVM research are highlighted sporadically in that section. Finally, Section \ref{conclusion} concludes the paper.  

\section{\projname}
\label{benchmark}
Benchmarks play an important role in evaluating the effectiveness and generalization when an architecture optimization is proposed. We develop a comprehensive UVM benchmark suite to facilitate the research on the GPU UVM. This suite covers a wide range of application domains marked in Table \ref{benchname}. The benchmarks exhibit diverse memory access patterns (more in Section \ref{pattern}) to help evaluate memory management strategies in GPU UVM. The suite also includes several auxiliary python-based programs to help create and test memory oversubscription cases. The benchmark suite is referred to as \textit{\projname}, and has been made available to the GPU research community for both non-UVM and UVM versions (\url{https://github.com/OSU-STARLAB/UVM_benchmark}). Table \ref{benchname} lists all the benchmarks and their configurations in UVMbench. Table \ref{bench_comp} compares the UVMbench with some related but limited workloads in several important aspects. 
The development of the benchmark suite includes the following major efforts.


\textbf{(1) Re-implement existing benchmarks.} We start with combining three existing popular GPU benchmark suites, i.e., Rodinia \cite{che2009rodinia, che2010characterization}, Parboil \cite{stratton2012parboil} and Polybench \cite{pouchet2012polybench}, removing redundant workloads and workload types, and converting into the UVM-based programming model. To implement UVM for these benchmarks, we replace all the host pointers (CPU side) and device pointers (GPU side) with a unified pointer allocated by the UVM API \textit{cudaMallocManaged}. Also, because the GPU driver is now responsible for data migration, all the explicit memory data migration APIs in each original program need to be removed. This may involve rewriting part of the code around the API calls in some benchmarks to achieve the equivalent functionalities. Moreover, the non-UVM data allocation structure should be adapted to the UVM version. For instance, we have to flatten non-UVM 2D arrays, previously allocated on the host side, into 1D arrays, as no 2D array allocation API is provided in the UVM programming model.

\textbf{(2) Develop machine learning workloads.} As recent machine learning tasks heavily rely on GPUs for acceleration, we also add more machine learning related workloads in our benchmark suite, as briefly described below:
\begin{itemize}
	\item \textit{Bayesian Network (BN)} is a probabilistic-based graphical model, often used for predicting the likelihood of several possible causes given the occurrence of an event. Our implementation is based on the SJTU version \cite{wang2015learning} and, during the conversion to UVM, retains the two phases that are accelerated by the GPU: preprocessing where local scores of every possible parent set for each node are calculated, and score calculation where threads obtain the local scores and return the best one. 
	
	\item \textit{Convolutional Neural Network (CNN)} is most commonly applied to image recognition. It has also been extended to video analysis, natural language processing and many other fields. Our implementation follows the general practice where, for forward propagation, the kernels of convolutional operations, activation operations and fully connected operations are accelerated on the GPU; and for back propagation, the kernels on error calculations and weight and bias update operations are accelerated on the GPU.
	
	
	\item \textit{Logistic Regression (LR)} is used to predict the probability of the existence of a certain class or event. The cost calculation is accelerated on the GPU. The input of this benchmark is the document-level sentiment polarity annotations which is first introduced in \cite{maas-EtAl}.


	\item \textit{Support Vector Machine (SVM)} is to find support vectors that, collectively, form a hyper plane to separate different classes. In our implementation, the kernel matrix calculation is accelerated on the GPU. The code is based on the Julia project \cite{qinyu} and converted to UVM.
	
\end{itemize}

Listings \ref{lst:1} and \ref{lst:2} show the partial code of the sigma update function in the SVM benchmark, which demonstrates the re-implementation process and newly added benchmarks. Several unrelated variables are omitted for simplicity. Listing \ref{lst:1} is the code without UVM, while Listing \ref{lst:2} is the code with UVM during runtime. As the traditional programming model requires explicit memory management, the program in Listing \ref{lst:1} has to allocate memory space on the device by calling \textit{CudaMalloc} (lines 12-20). It also needs to call \textit{CudaMemcpy} APIs (lines 22-24 and lines 28-31) before and after the kernel launch to explicitly migrate the required data between the host and the device. In contrast, the UVM programming model in Listing \ref{lst:2} unifies the memory space of the host and the device. By calling \textit{cudaMallocManaged} APIs (lines 6-7), the code allocates bytes of managed memory. The allocated variables can be accessed by the host and the device directly, and are managed by the Unified Memory system of the GPU. In Listing \ref{lst:2}, when this Sigma\_update function is called in the main function (line 1), the variables, defined by \textit{cudaMallocManaged}, are passed into the function, and the device kennels can directly access these variables. Therefore, device variable definitions and memory management APIs are removed (i.e., lines 6-7 in Listing \ref{lst:2} vs. lines 12-20 \& 22-24 \& 28-31 in Listing \ref{lst:1}). It can be seen that the UVM programming model greatly reduces the code complexity.

\begin{lstlisting}[caption={\textit{Sigma\_Update} function in SVM with non-UVM.},label={lst:1}]
Sigma_update(int *iters, float *alpha, float *sigma,float *K, int *y, int l, int C)
{
//Define variables on the device
float *dev_alpha = 0;
float *dev_sigma = 0;
float *dev_K = 0;
int *dev_y = 0;
int *dev_block_done = 0;
float *dev_delta = 0;
void *args[10] = {&dev_iters, &dev_alpha, &dev_sigma, &dev_K, &dev_y, &dev_block_done, &grid_dimension, &dev_delta, &l, &C};

//Allocate memory space on the device memory 
cudaMalloc(&dev_iters, sizeof(int));
cudaMalloc(&dev_alpha, l*sizeof(float));
cudaMalloc(&dev_sigma, l*sizeof(float));
cudaMalloc(&dev_K, l*l*sizeof(float));
cudaMalloc(&dev_y, l*sizeof(int));
cudaMalloc(&dev_block_done, 
grid_dimension*sizeof(int));
cudaMalloc(&dev_delta, 1*sizeof(float));

//Data migration: Host to Device
cudaMemcpy(dev_K, K, l*l*sizeof(float), cudaMemcpyHostToDevice);
cudaMemcpy(dev_y, y, l*sizeof(int), cudaMemcpyHostToDevice);

/*Kernel Launch*/

//Data migration: Device to Host
cudaMemcpy(iters, dev_iters, sizeof(int), cudaMemcpyDeviceToHost);
cudaMemcpy(alpha, dev_alpha, l* sizeof(float), cudaMemcpyDeviceToHost);
cudaMemcpy(sigma, dev_sigma, l* sizeof(float), cudaMemcpyDeviceToHost);

//Free allocated memory space
cudaFree(dev_block_done);
cudaFree(dev_delta);
cudaFree(dev_y);
cudaFree(dev_K);
cudaFree(dev_sigma);
cudaFree(dev_alpha);
cudaFree(dev_iters);
}
\end{lstlisting}

\begin{lstlisting}[caption={\textit{Sigma\_Update} function in SVM with UVM.},label={lst:2}]
Sigma_update(int *iters, float *alpha, float *sigma, float *K, int *y, int l, int C)
{
int *dev_block_done = 0;
float *dev_delta = 0;
void *args[10] = {&iters, &alpha, &sigma, &K, &y, &dev_block_done, &grid_dimension, &dev_delta, &l, &C};
cudaMallocManaged(&dev_block_done, grid_dimension*sizeof(int));
cudaMallocManaged(&dev_delta, 1*sizeof(float));

/*Kernel Launch*/

cudaFree(dev_block_done);
cudaFree(dev_delta);
}
\end{lstlisting}

\textbf{(3) Optimize data prefetch.} In our experiment, we observe that directly converting to the UVM programing model from the non-UVM model can lead to performance degradation, as UVM has to track memory accesses and migrate data to destinations. Therefore, we add an optimization, namely asynchronous prefetching, before each kernel launch by calling the provided API \textit{cudaMemPrefetchAsync}. The purpose of this optimization is to exemplify that hardware prefetchers may bring considerable performance improvement in UVM, as shown later in evaluation results. Users of our benchmark suite can easily enable or disable this optimization by changing the macro definition in the Makefile. 

Listing \ref{lst:3} shows the code in the Backprop benchmark after enabling the above asynchronous prefetching. The program uses CUDA streams to manage concurrency in GPU applications. Different streams can execute their corresponding commands concurrently. To prepare asynchronous prefetching, it first creates different streams (lines 2-11). With different streams, the prefetching APIs (lines 13-16 and 23-24) prefetch the required data asynchronously. As the data have been fetched in the device before the kernel is launched, the Unified Memory system does not need to stall the kernel and handle page faults. Therefore, the data migration overhead in the UVM is mitigated under asynchronous prefetching. 

\begin{lstlisting}[caption={Enable Prefetching in \textit{Backprop} with UVM.},label={lst:3}]
//Create streams for asynchronous prefetch
cudaStream_t stream1;
cudaStream_t stream2;
cudaStream_t stream3;
cudaStream_t stream4;
cudaStream_t stream5;
cudaStreamCreate(&stream1);
cudaStreamCreate(&stream2);
cudaStreamCreate(&stream3);
cudaStreamCreate(&stream4);
cudaStreamCreate(&stream5);

cudaMemPrefetchAsync(input_cuda, (in + 1)*sizeof(float), 0, stream1);
cudaMemPrefetchAsync(output_hidden_cuda, (hid + 1)*sizeof(float), 0, stream2);
cudaMemPrefetchAsync(input_hidden_cuda, (in + 1)*(hid + 1)*sizeof(float), 0, stream3);
cudaMemPrefetchAsync(hidden_partial_sum, num_blocks*WIDTH*sizeof(float), 0, stream4);

//Performing GPU computation

bpnn_layerforward_CUDA<<<grid, threads, 0, stream5>>>(input_cuda, output_hidden_cuda, input_hidden_cuda, hidden_partial_sum, in, hid);
cudaDeviceSynchronize();

cudaMemPrefetchAsync(input_prev_weights_cuda, (in + 1)*(hid + 1) sizeof(float), 0, stream1);
cudaMemPrefetchAsync(hidden_delta_cuda, (hid + 1)*sizeof(float), 0, stream2);

bpnn_adjust_weights_cuda<<<grid, threads, 0, stream5>>>(hidden_delta_cuda, hid, input_cuda, in, input_hidden_cuda, input_prev_weights_cuda);
cudaDeviceSynchronize();

\end{lstlisting}
\textbf{(4) Optimize data reuse.} Data reuse can also mitigate performance overhead of UVM. This is because if the useful data resides in the device memory for longer time, fewer page faults may occur. To investigate the impact of data reuse where multiple (same) kernels access the same data during the runtime, we add the option to run multiple iterations of a kernel execution to create this type of data reuse opportunities (i.e., the same kernel reuses the same data in different iterations). Users can change the number of iterations ($\ge 1$) by modifying the macro in each benchmark program file. 


Benchmarks in the proposed \projname \ are all implemented in CUDA and can be run on Nvidia GPUs. This suite includes both the non-UVM version (original) and the UVM version implementation for performance comparison. There are no algorithmic changes when developing the UVM version of the benchmarks. This ensures fair comparison between the traditional programming model and the UVM programming model. Consequently, the observed performance changes are mostly attributed to the difference between programming models rather the algorithms.

Some previous works \cite{ganguly2019interplay,chien2019performance} and the Nvidia SDK present a limited number of UVM-enabled workloads to demonstrate the effectiveness of the UVM or their proposed ideas. Table \ref{bench_comp} compares the existing benchmarks with our proposed UVMbench in five important aspects. Compared with the existing benchmarks, UVMbench presents more workloads from different domains. In particular, UVMbench includes machine learning workloads to explore the possibility of applying UVM techniques in data-intensive machine learning applications. Moreover, UVMbench provides diverse memory access patterns and supports memory oversubscription.

\section{Evaluation Methodology}
\label{methodology}

Our evaluation methodology is designed to enable a set of experiments that test the proposed benchmark suite. To investigate the impact of memory access behaviors on UVM, we need to profile memory access patterns of each benchmark. Direct performance comparison is also needed between the UVM and non-UVM implementations. As the driver is responsible for data migration under UVM, the impact on PCIe bandwidth should also be examined. Additional experiment is needed to evaluate the UVM performance under memory oversubscription scenarios.

To conduct the above experiments, we employ an Nvidia GTX 1080 Ti GPU with the Pascal architecture. We use the Nvidia Binary Instrumentation Tool (NVBit) \cite{villa2019nvbit} to extract the global memory access patterns of the \projname \ suite. NVBit provides a fast, dynamic and portable binary instrumentation framework that allows users to inspect/instrument instructions. we use two Nvidia official profiling tools to profile the performance related data of benchmarks: nvprof, a command line tool to collect and view profiling data, and Nvidia Visual Profiler, a GUI to visualize the application performance. Table \ref{t2} includes more details of the CPU-GPU platform.



\begin{table}[t]
	\centering
	\small
	\def\arraystretch{1.3}
	\caption{Evaluation Platform Setup.}
	\begin{tabular}{cc}
		\thickhline
		CPU & Intel Xeon E5-2630 V4 10 Cores 2.2 GHz \\
		Memory & DDR4 16GB x 4 \\
		PCIe & PCIe Gen3x16 16GB/s \\
		Operating System & Ubuntu 18.04 64bit \\ \hline
		GPU & Nvidia GTX1080Ti \\
		Driver version & 440.33.01 \\
		CUDA & CUDA 10.2 \\
		Profiling Tools & nvprof, Nvidia Visual Profiler, NVBit \\ \thickhline
	\end{tabular}
	\label{t2}
\end{table}

\section{Results and Analysis}
\label{results}

\begin{figure*}[bhpt]
	\centering
	\subfloat[2DCONV (R)]{
		\includegraphics[width=0.24\textwidth]{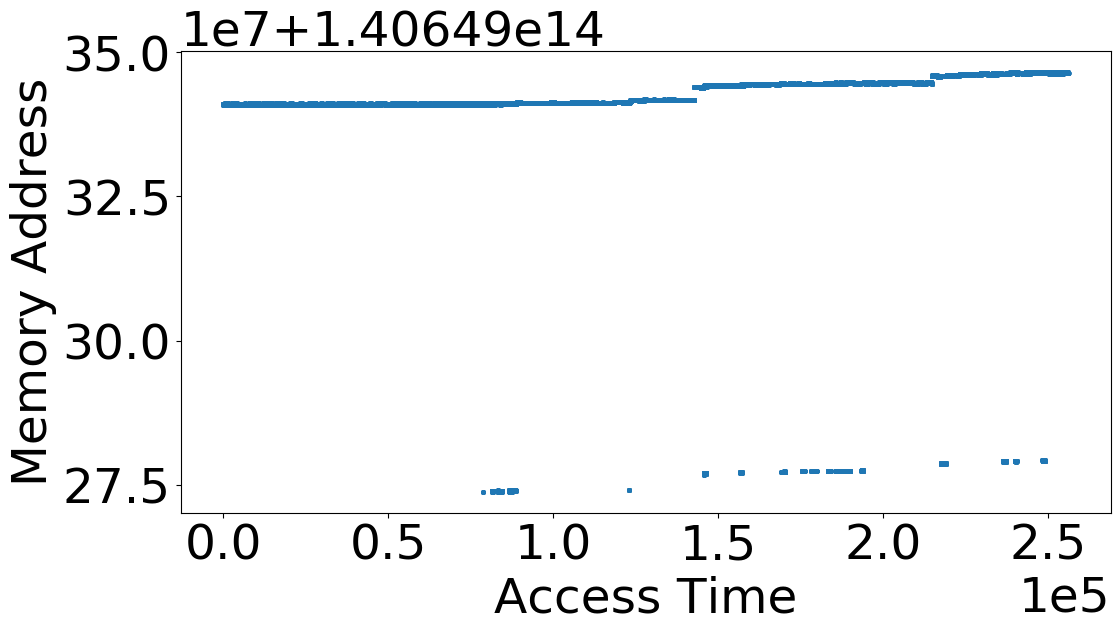}
	}
	\subfloat[2MM (R)]{
		\includegraphics[width=0.24\textwidth]{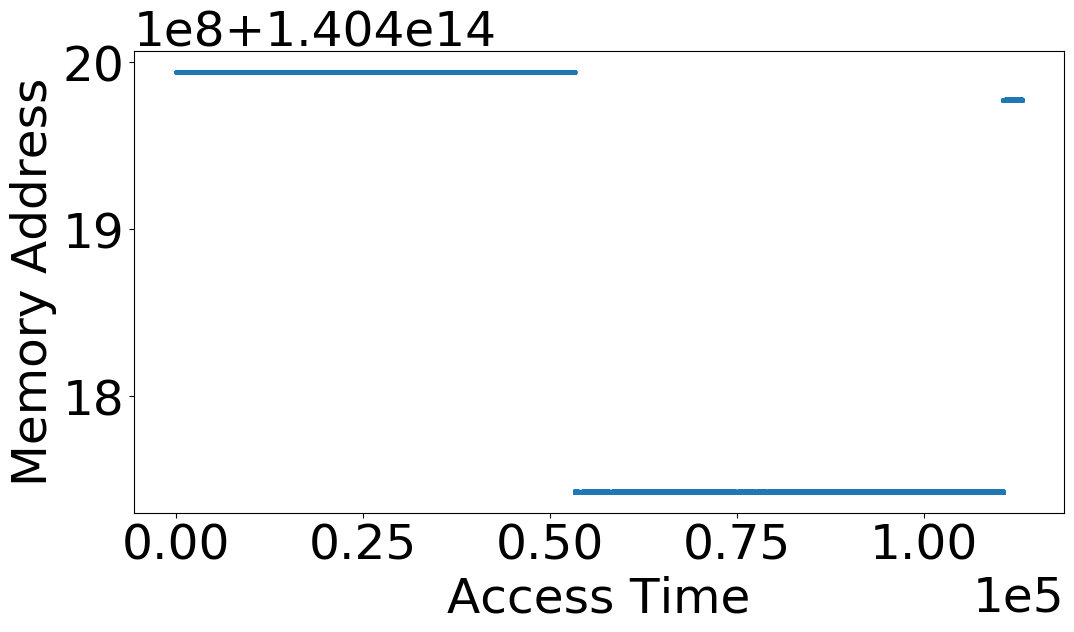}
	}
	\subfloat[3DCONV (R)]{
		\includegraphics[width=0.24\textwidth]{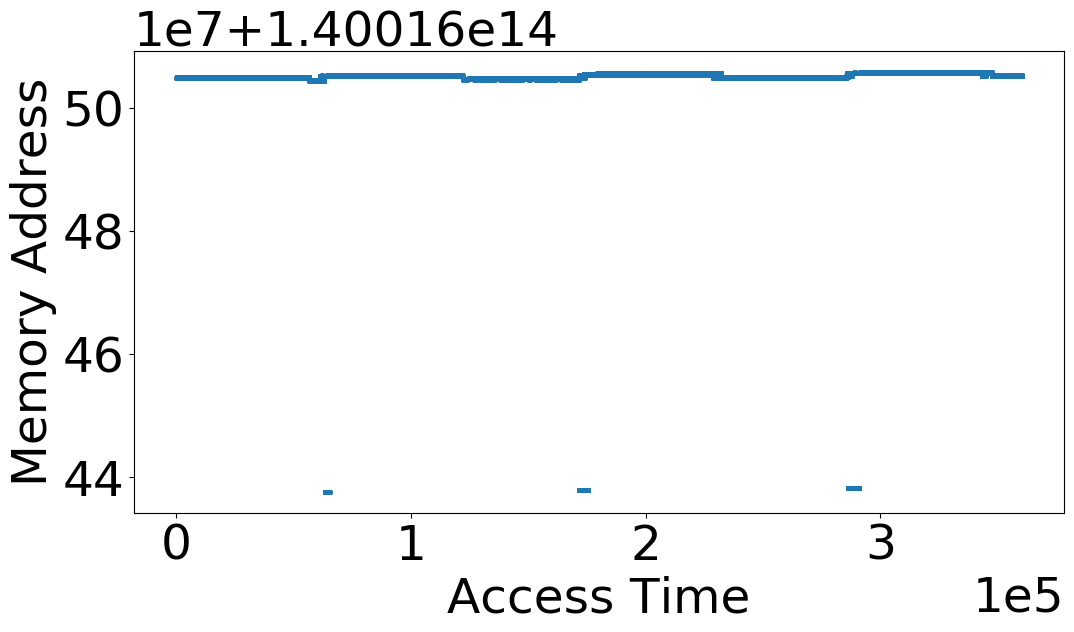}
	}
	\subfloat[3MM (R)]{
		\includegraphics[width=0.24\textwidth]{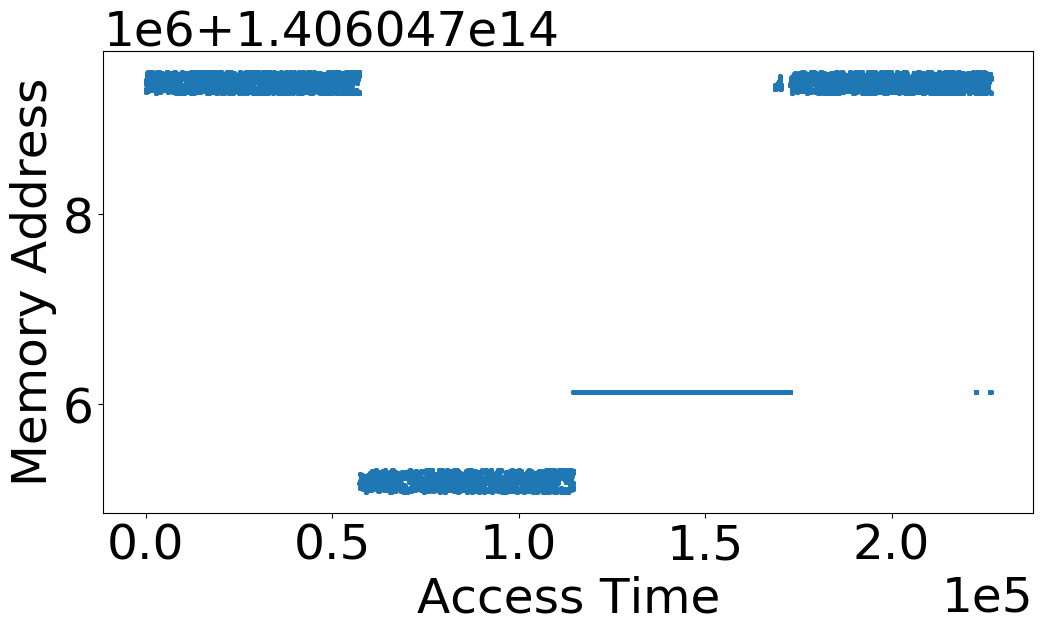}
	}
	\\
	\subfloat[ATAX (I)]{
		\includegraphics[width=0.24\textwidth]{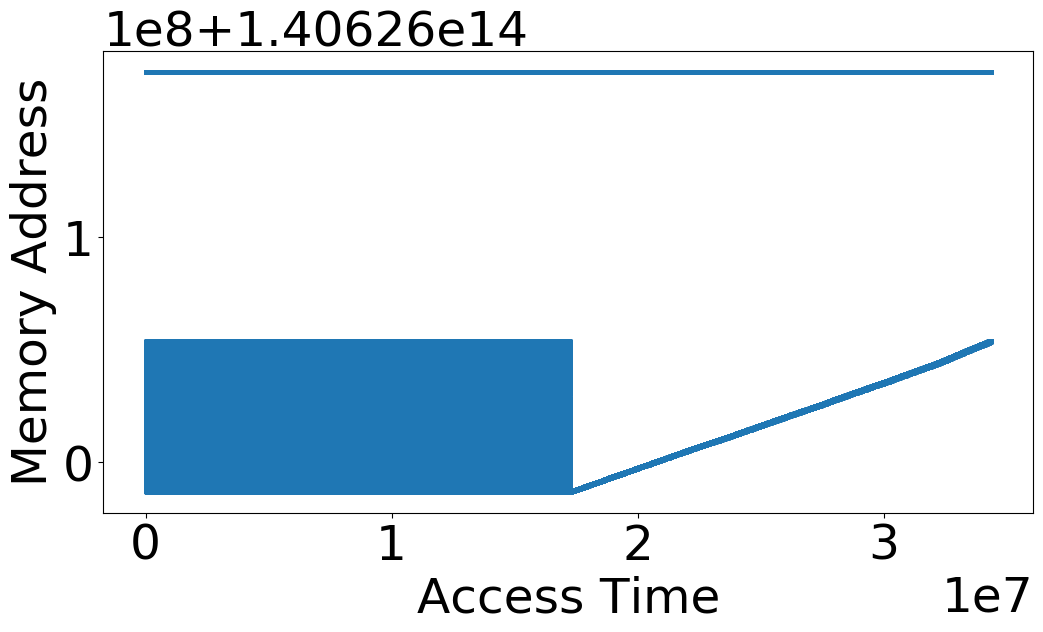}
	}
	\subfloat[BACKPROP (R)]{
		\includegraphics[width=0.24\textwidth]{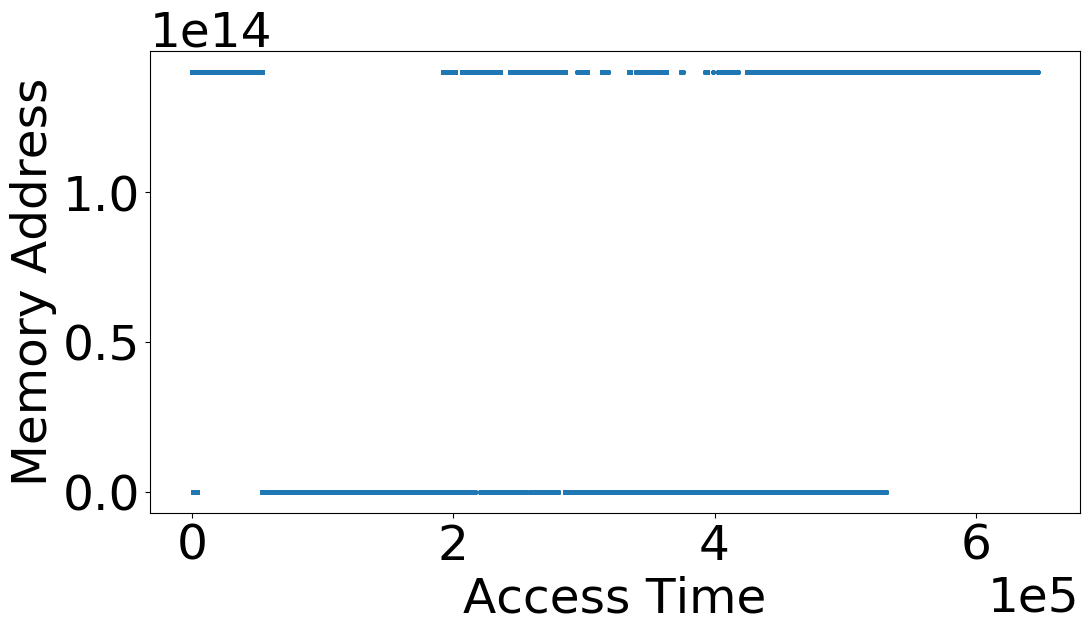}
	}
	\subfloat[BFS (I)]{
		\includegraphics[width=0.24\textwidth]{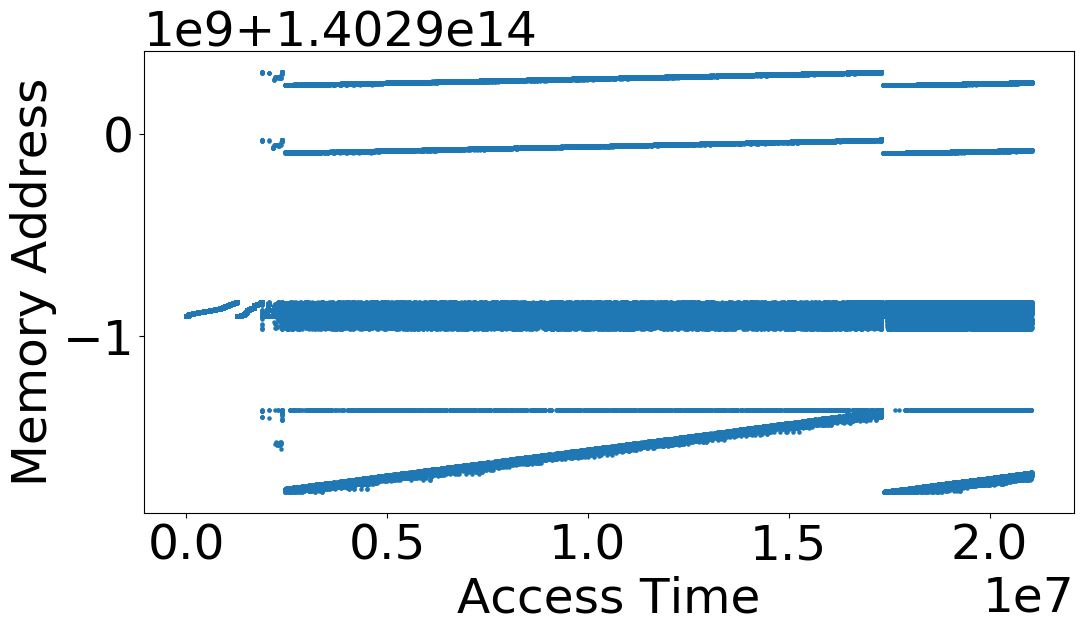}
	}
	\subfloat[BICG (I)]{
		\includegraphics[width=0.24\textwidth]{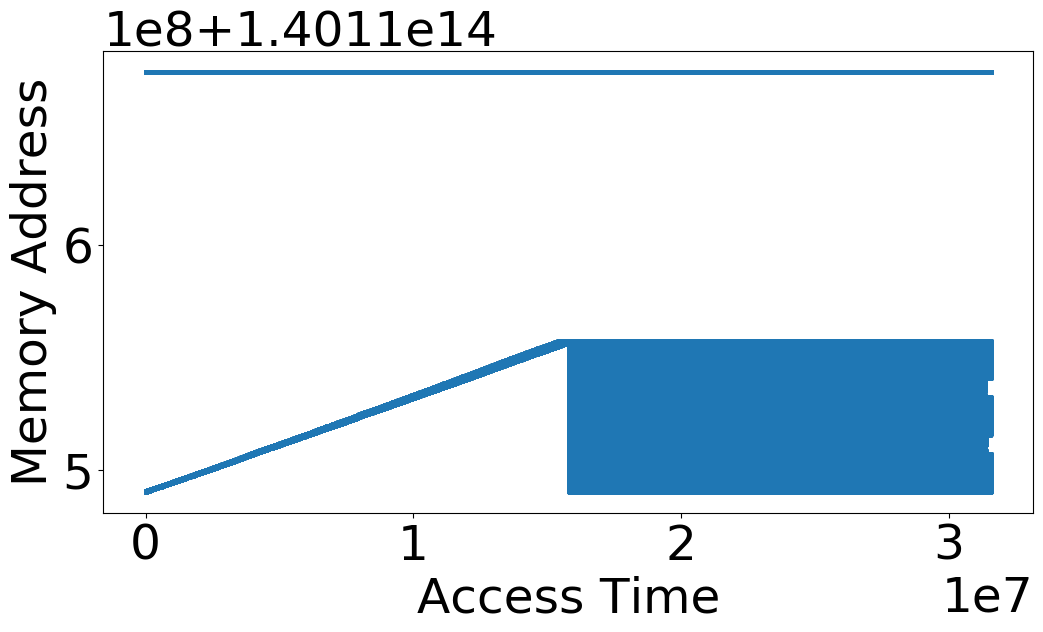}
	}
	\\
	\subfloat[BN (R)]{
		\includegraphics[width=0.24\textwidth]{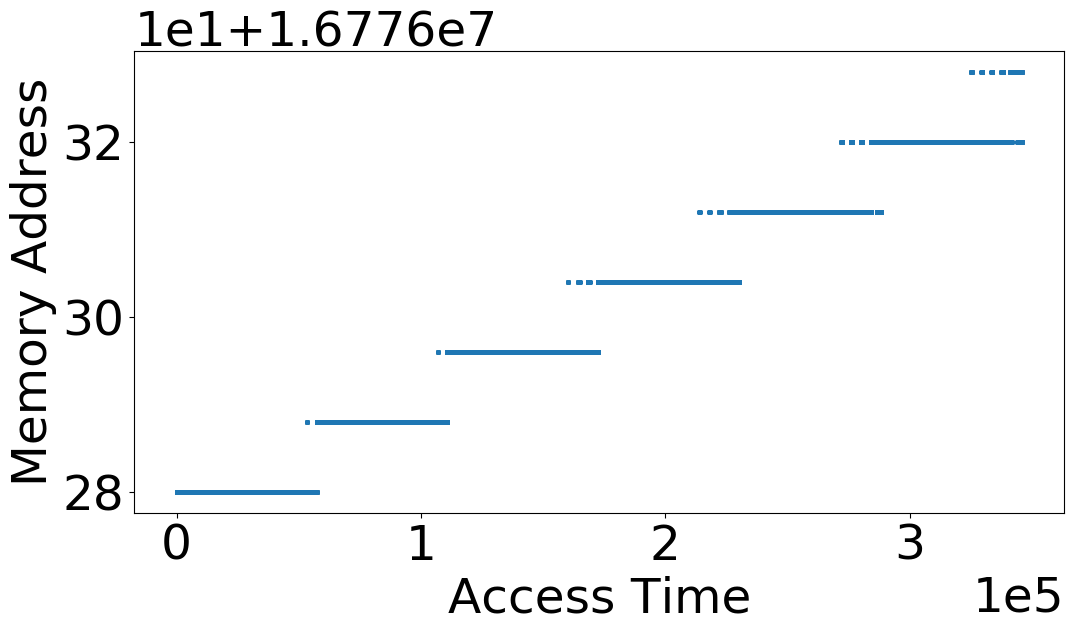}
	}
	\subfloat[CNN (R)]{
		\includegraphics[width=0.24\textwidth]{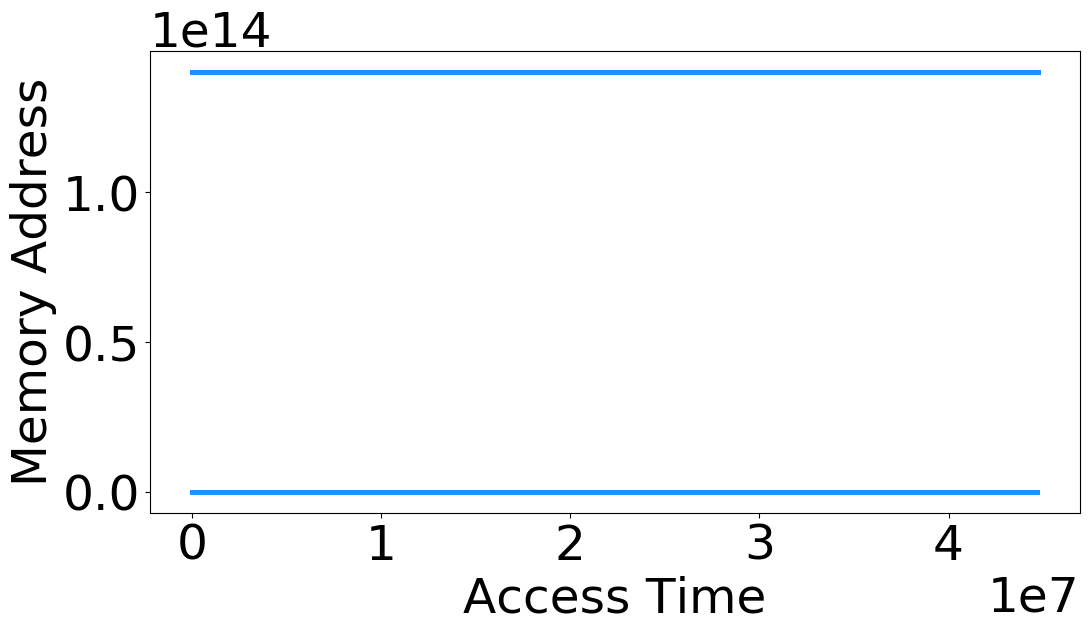}
	}
	\subfloat[CORR (I)]{
		\includegraphics[width=0.24\textwidth]{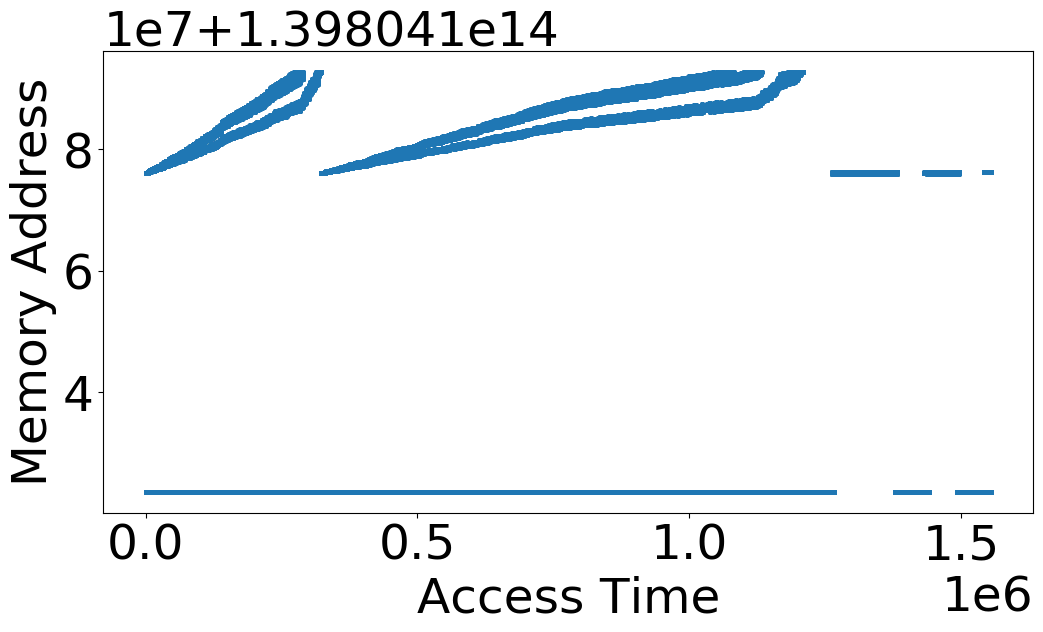}
	}
	\subfloat[COVAR (I)]{
		\includegraphics[width=0.24\textwidth]{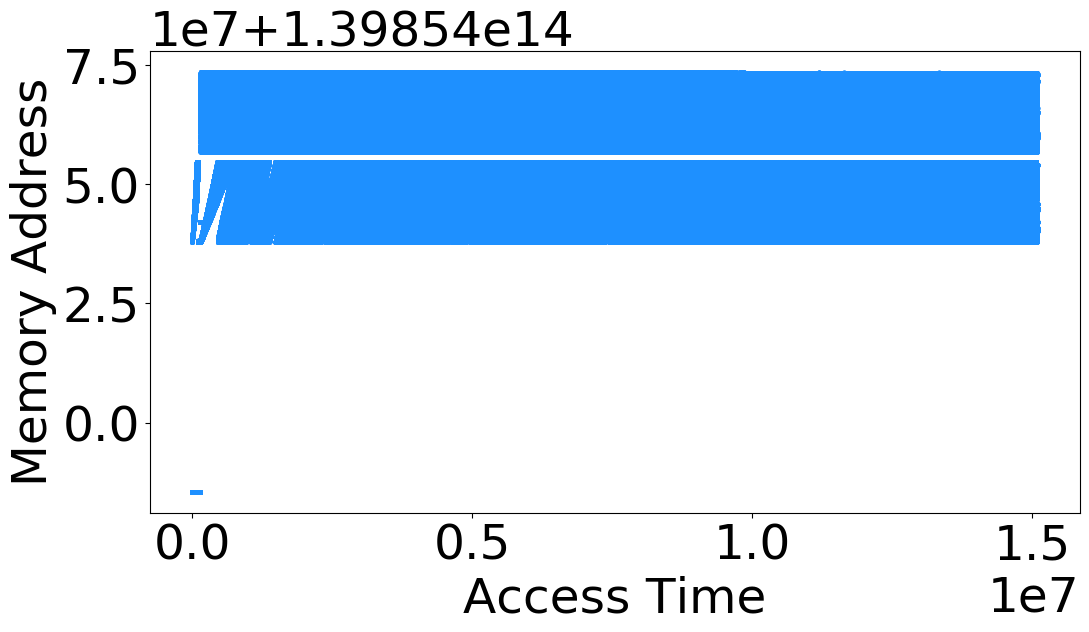}
	}
	\\
	\subfloat[DWT2D (R)]{
		\includegraphics[width=0.24\textwidth]{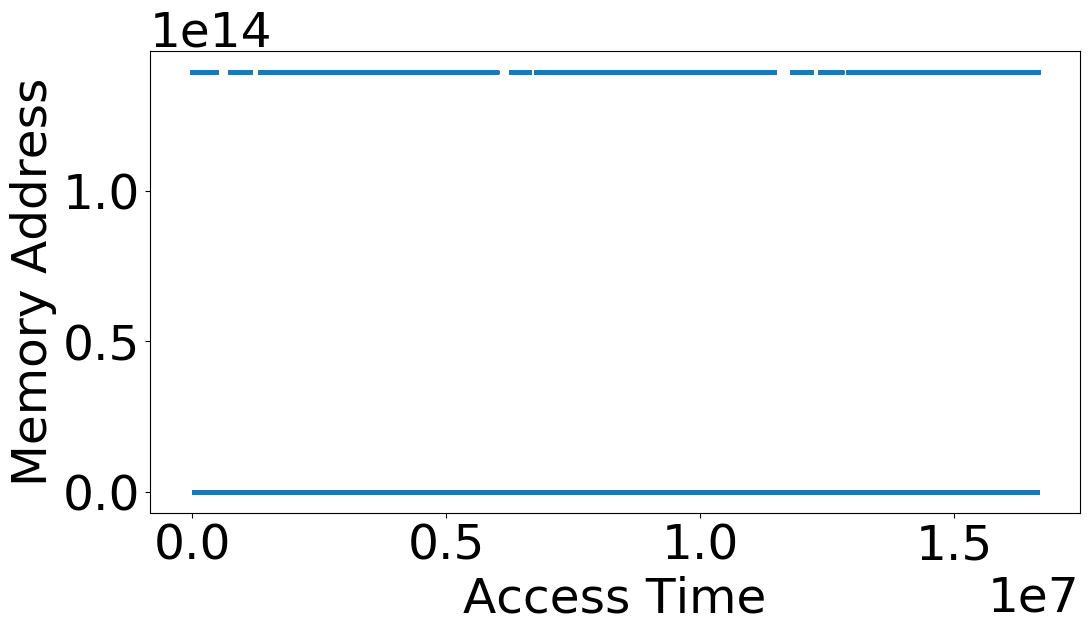}
	}
	\subfloat[FDTD-2D (I)]{
		\includegraphics[width=0.24\textwidth]{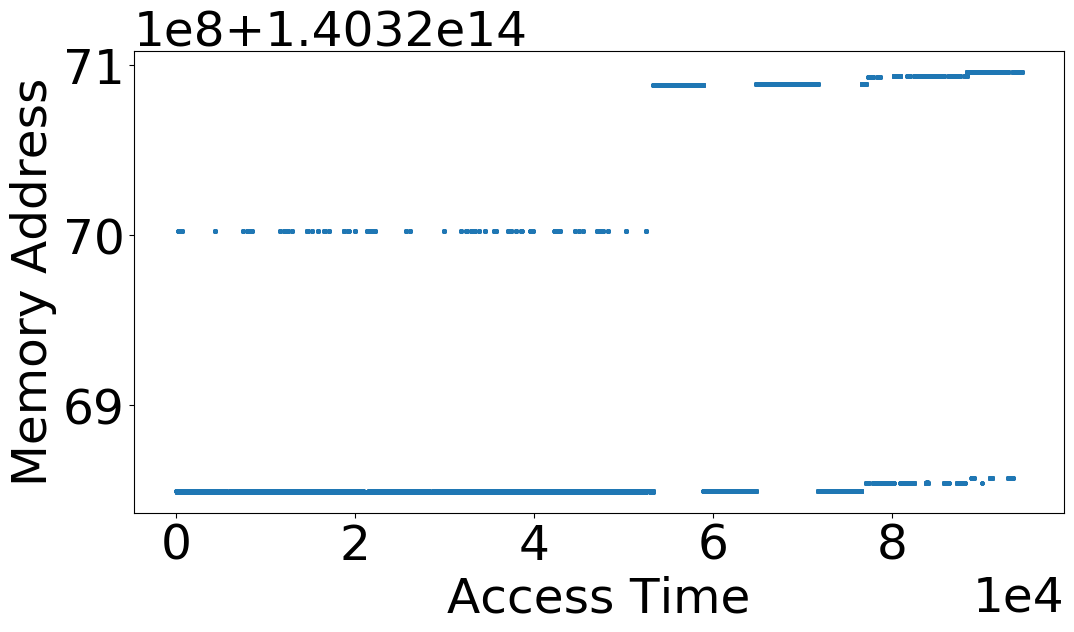}
	}
	\subfloat[GAUSSIAN (I)]{
		\includegraphics[width=0.24\textwidth]{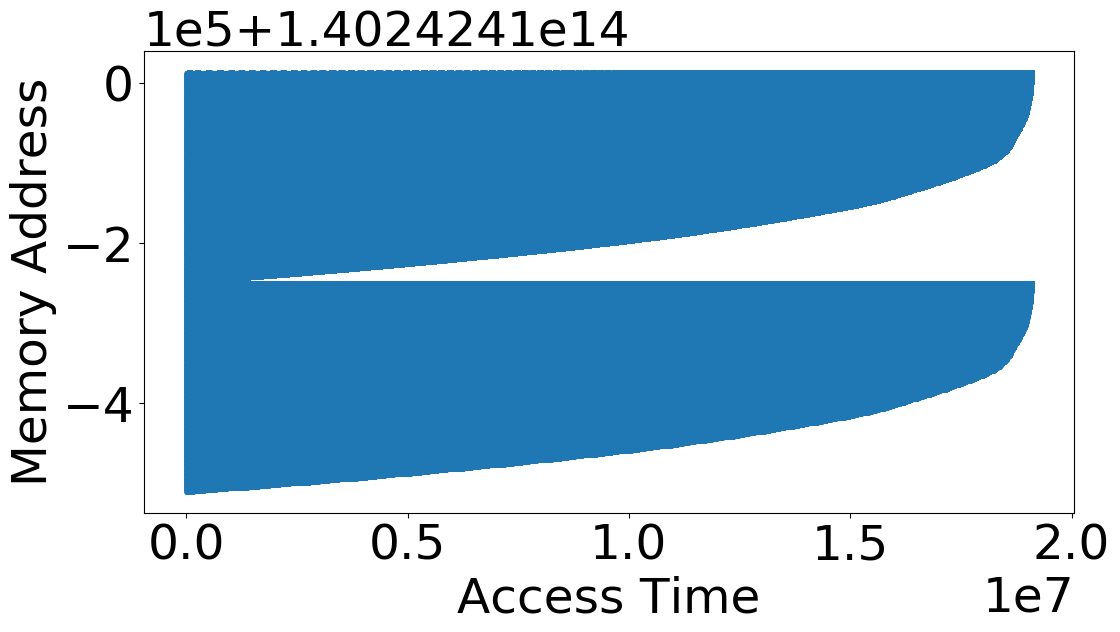}
	}
	\subfloat[GEMM (I)]{
		\includegraphics[width=0.24\textwidth]{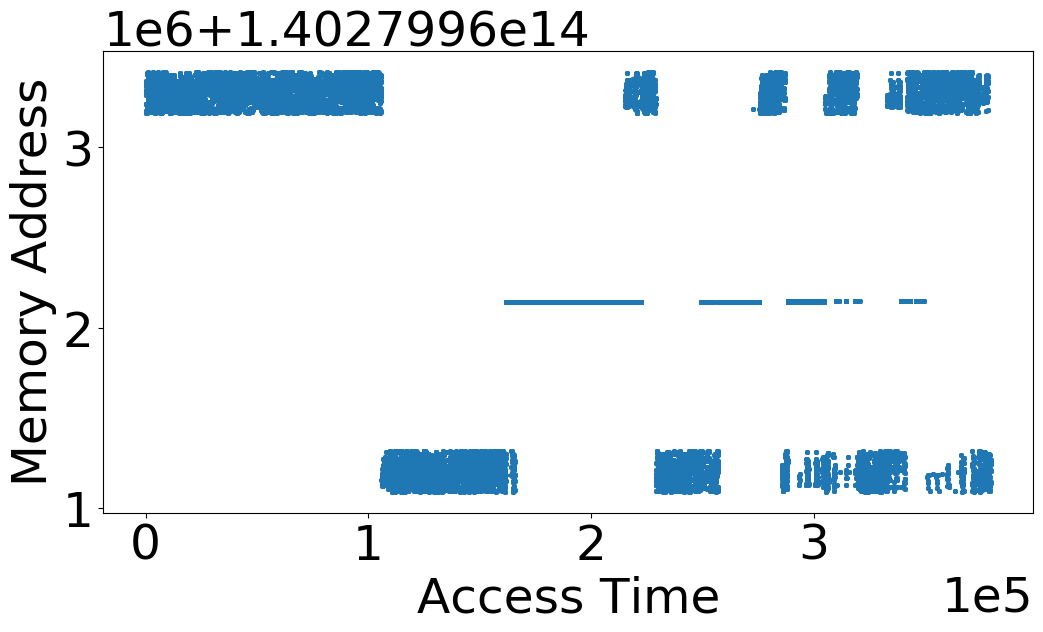}
	}
	\\
	\subfloat[GESUMMV (I)]{
		\includegraphics[width=0.24\textwidth]{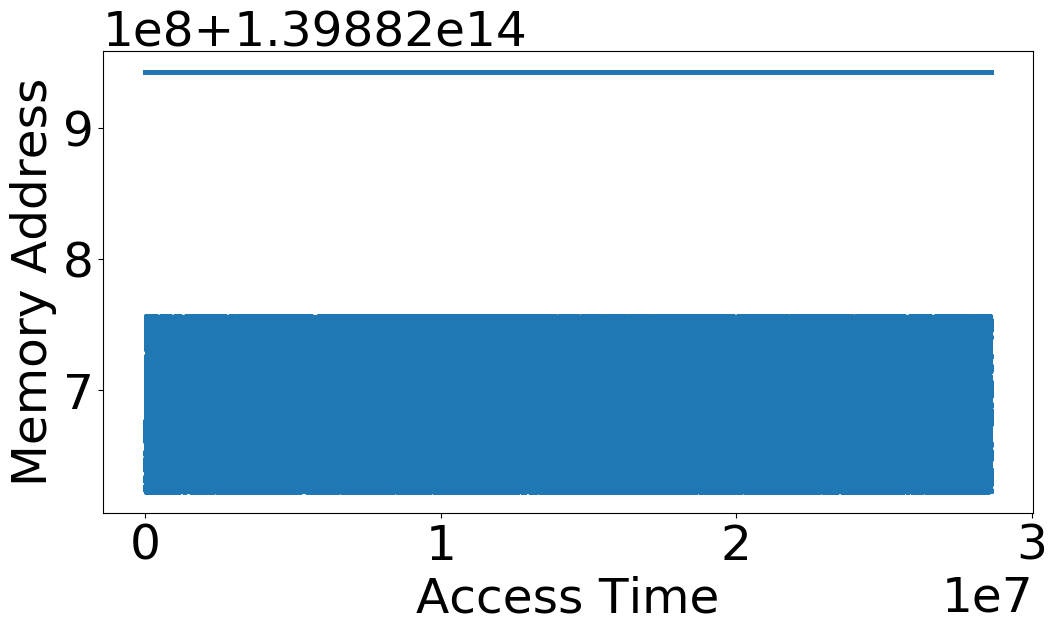}
	}
	\subfloat[GRAMSCHM (I)]{
		\includegraphics[width=0.24\textwidth]{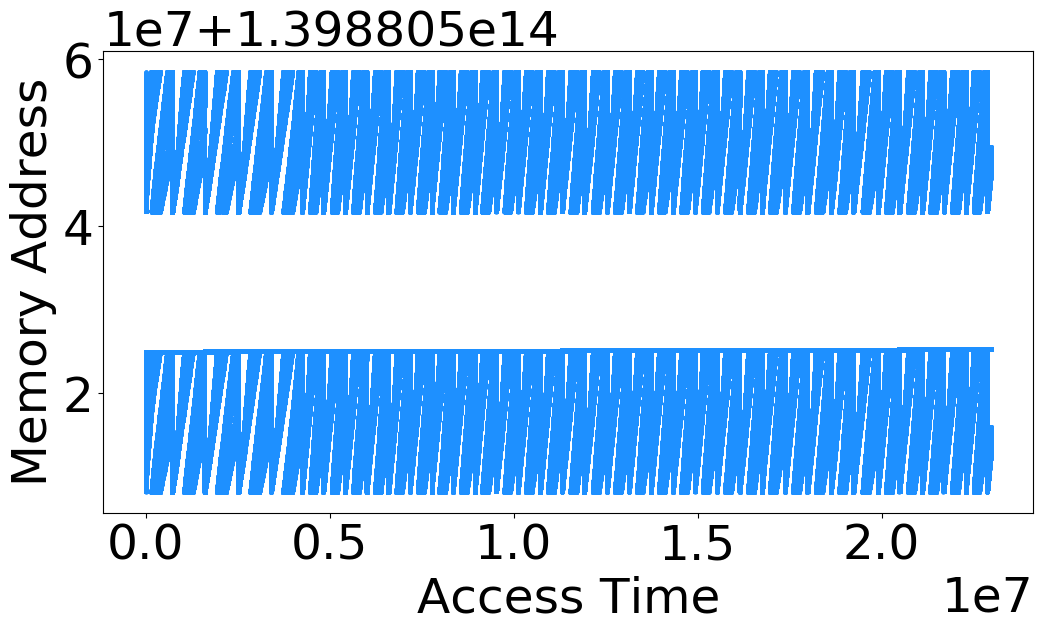}
	}
	\subfloat[HOTSPOT (R)]{
		\includegraphics[width=0.24\textwidth]{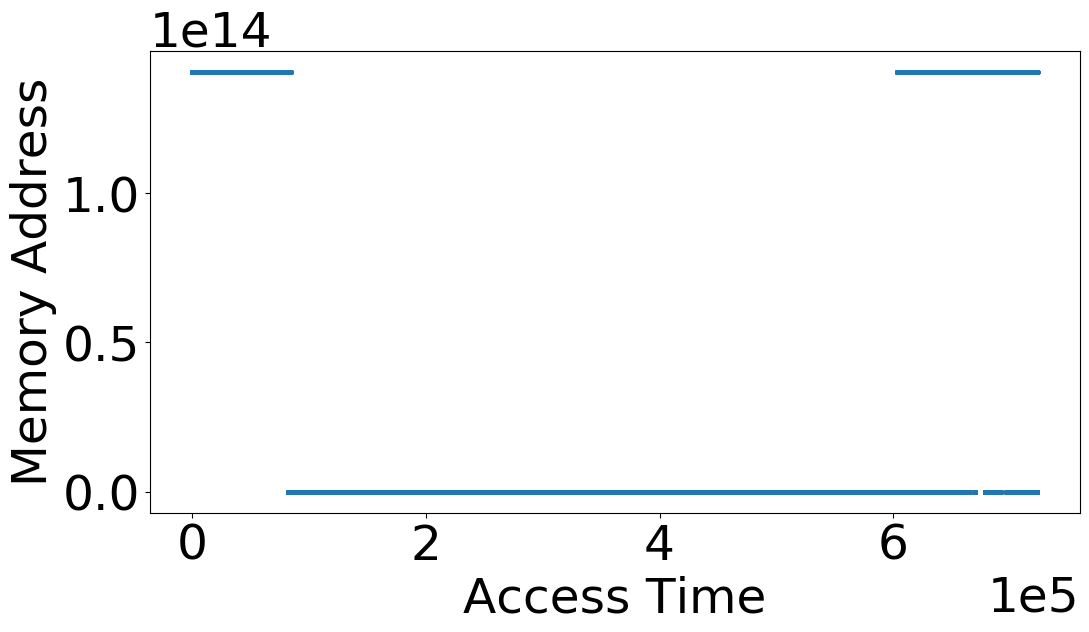}
	}
	\subfloat[HOTSPOT3D (R)]{
		\includegraphics[width=0.24\textwidth]{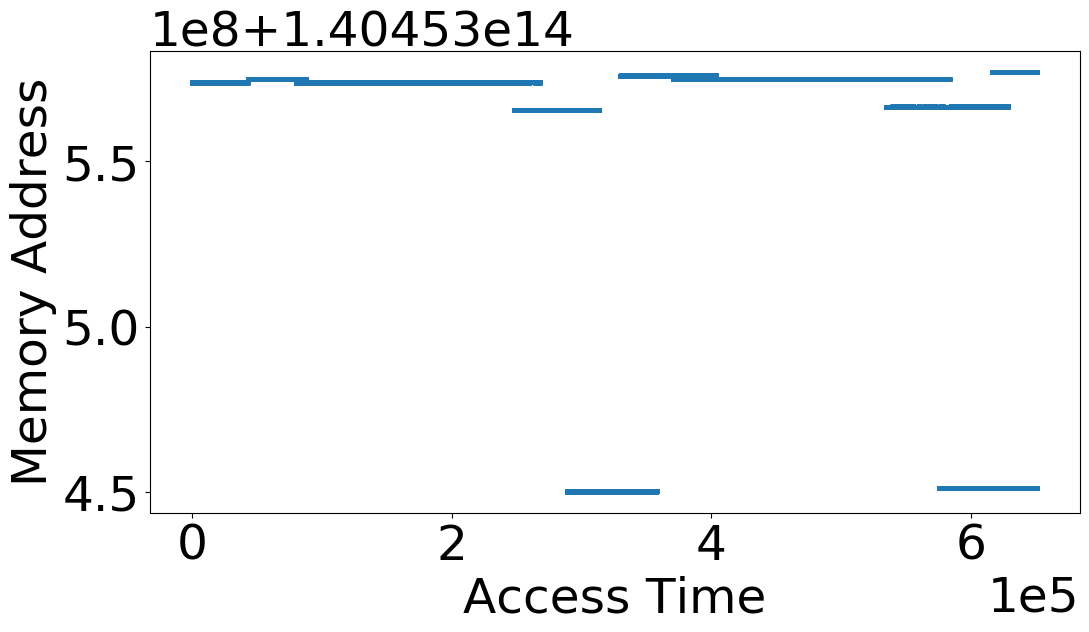}
	}
	\\
	\subfloat[KMEANS (I)]{
		\includegraphics[width=0.24\textwidth]{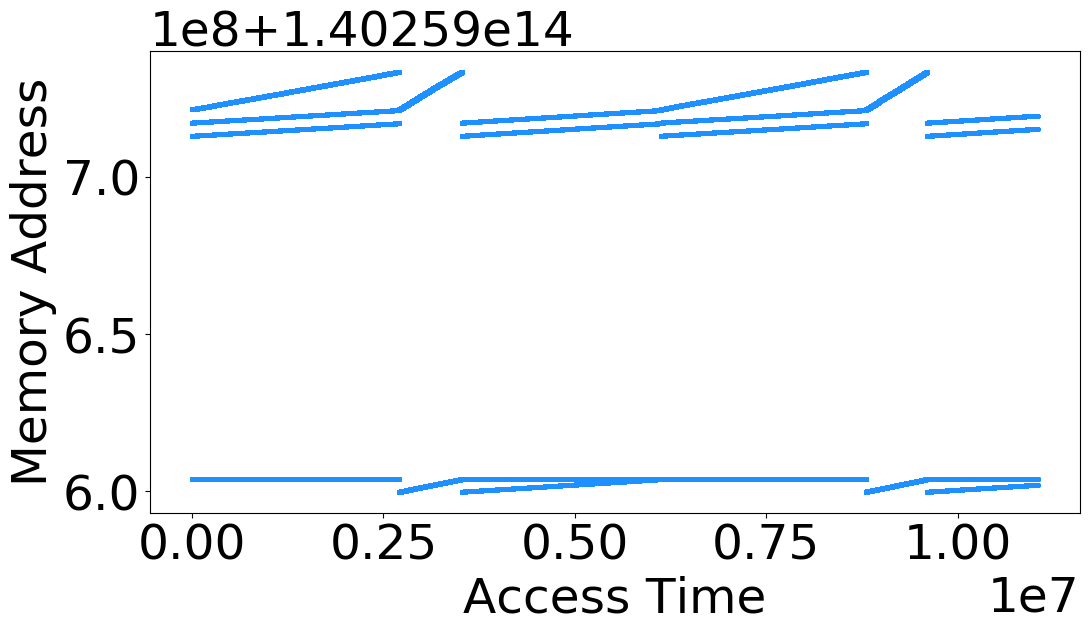}
	}
	\subfloat[KNN (R)]{
		\includegraphics[width=0.24\textwidth]{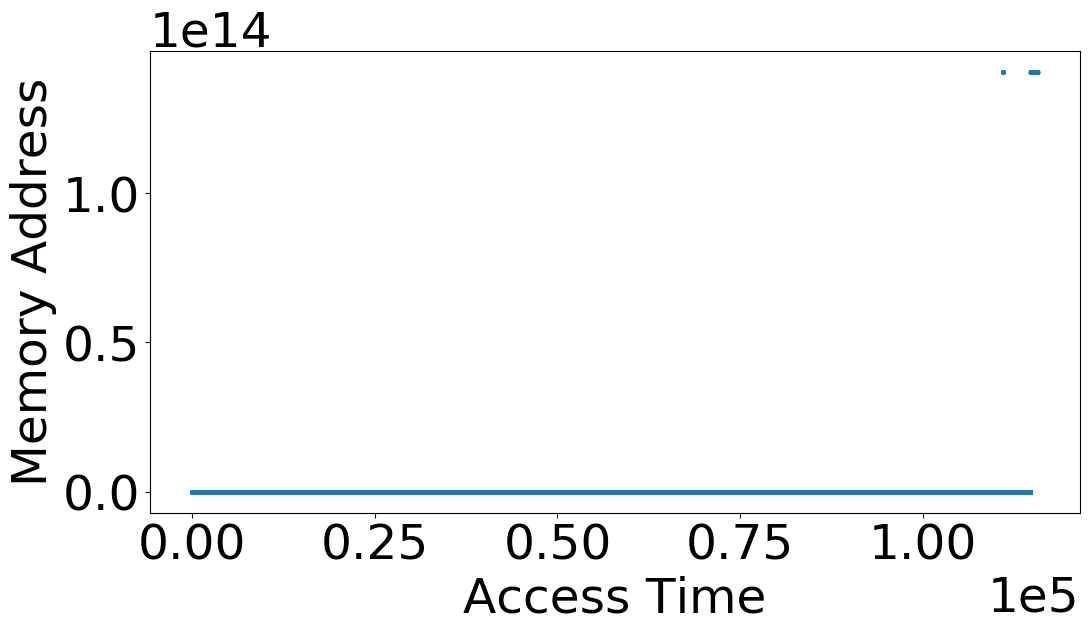}
	}
	\subfloat[LR (R)]{
		\includegraphics[width=0.24\textwidth]{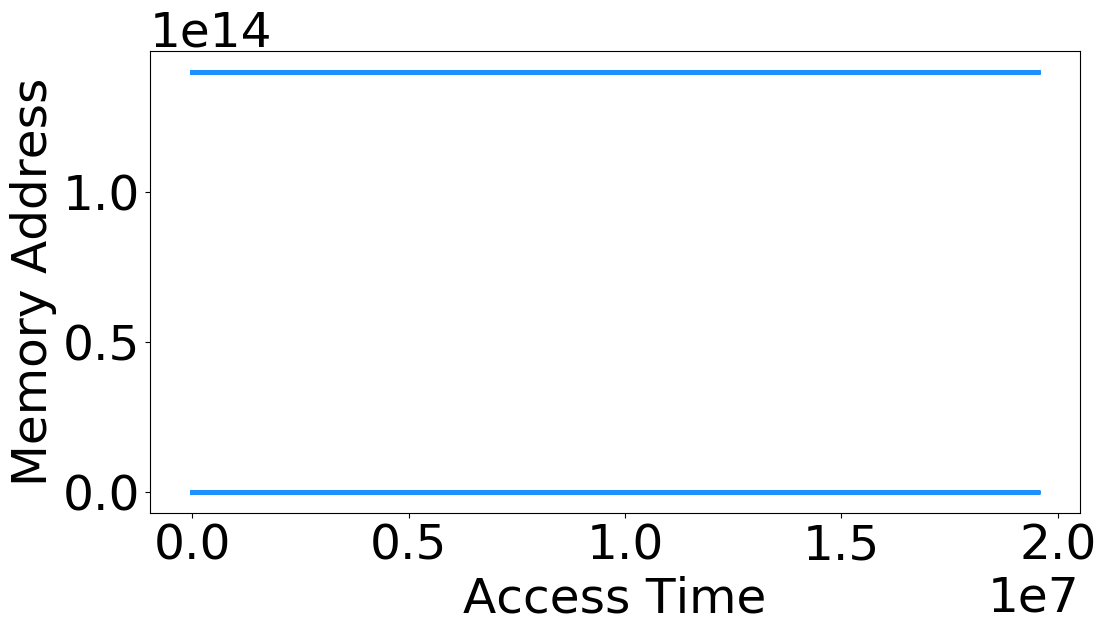}
	}
	\subfloat[MVT (I)]{
		\includegraphics[width=0.24\textwidth]{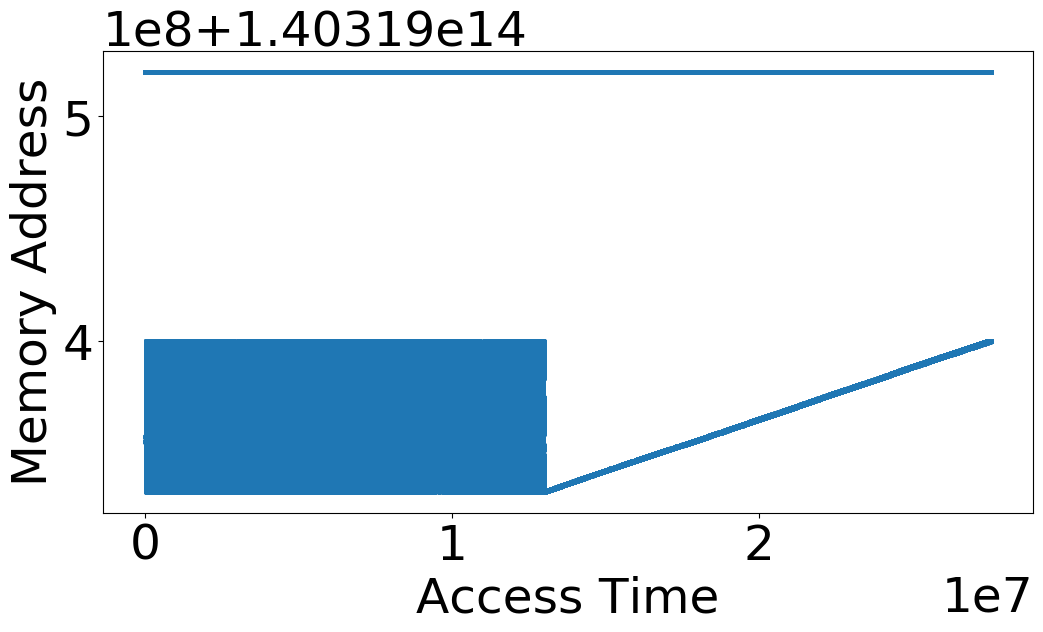}
	}
	\clearpage 
	\vspace{2ex}
	\caption{Memory access patterns of benchmarks in UVMBench.}
	\label{access_pattern}
\end{figure*}

\begin{figure*}[bhpt]\ContinuedFloat
	\centering
	\subfloat[NW (I)]{
		\includegraphics[width=0.24\textwidth]{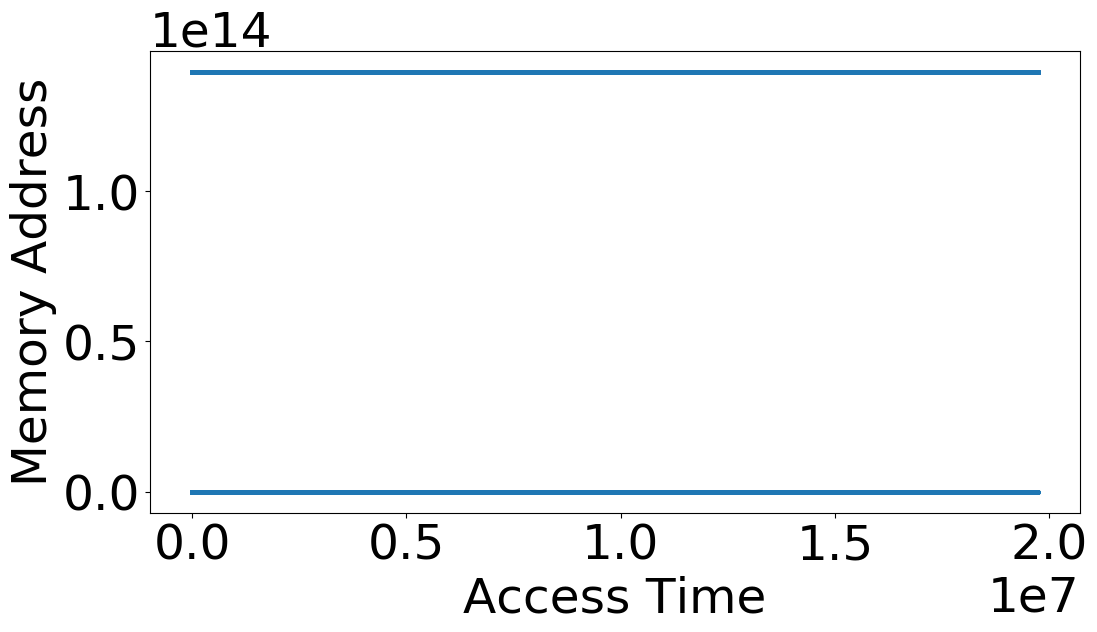}
	}
	\subfloat[PFILTER (R)]{
		\includegraphics[width=0.24\textwidth]{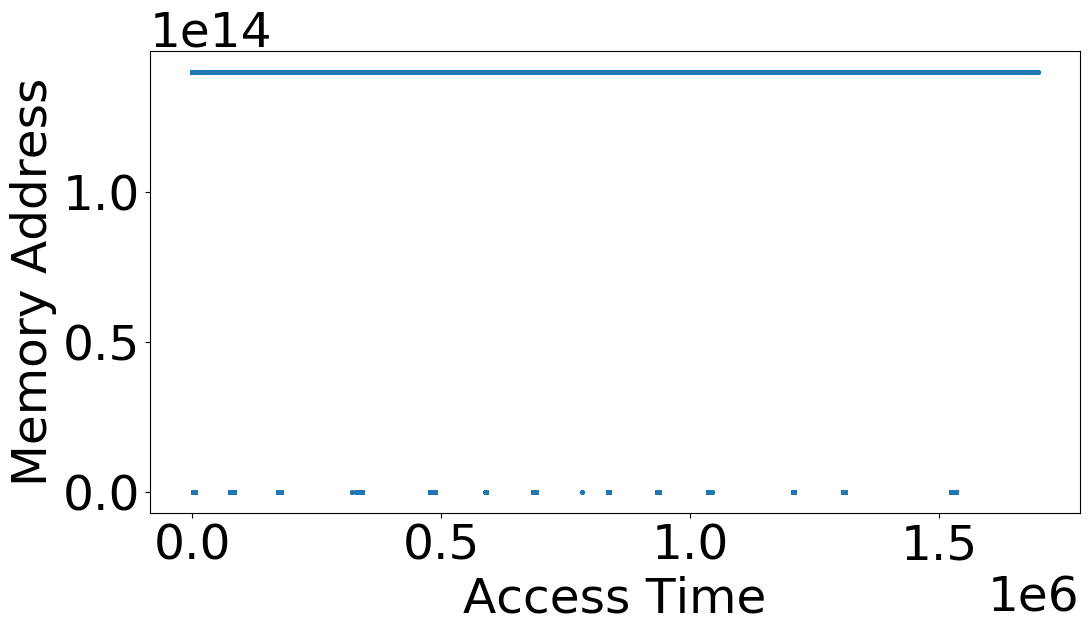}
	}
	\subfloat[PATHFINDER (R)]{
		\includegraphics[width=0.24\textwidth]{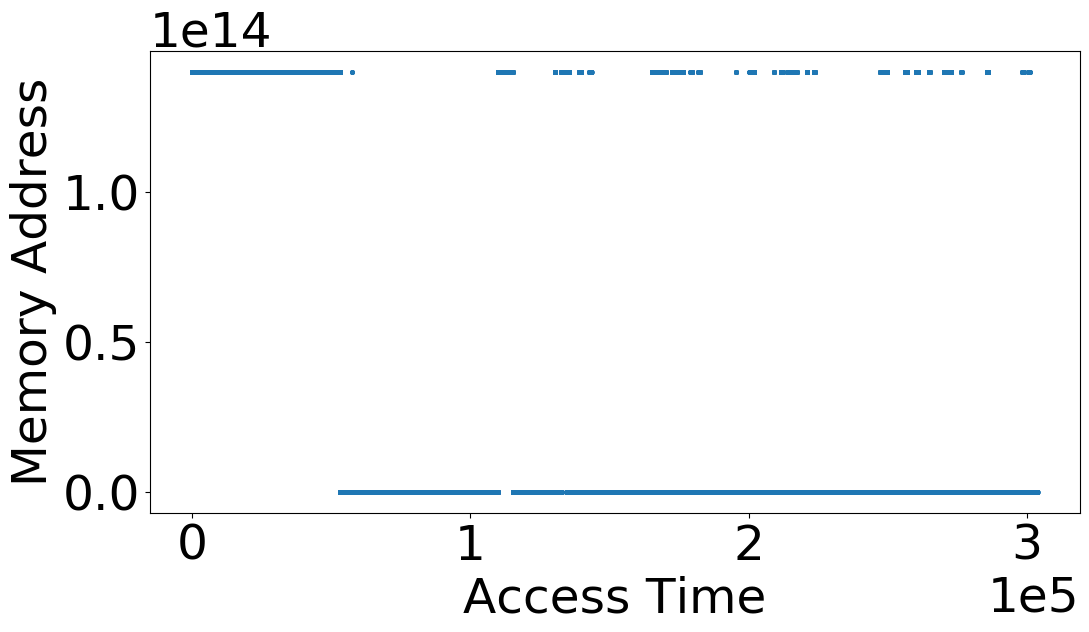}
	}
	\subfloat[SRAD (R)]{
		\includegraphics[width=0.24\textwidth]{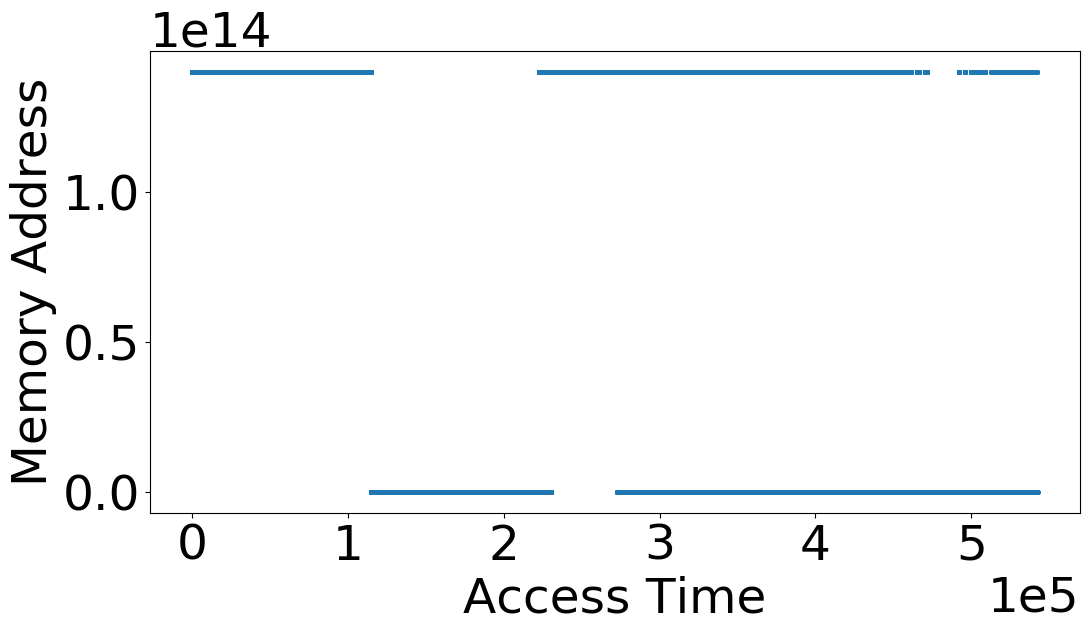}
	}
	\\
	\subfloat[SC (I)]{
		\includegraphics[width=0.24\textwidth]{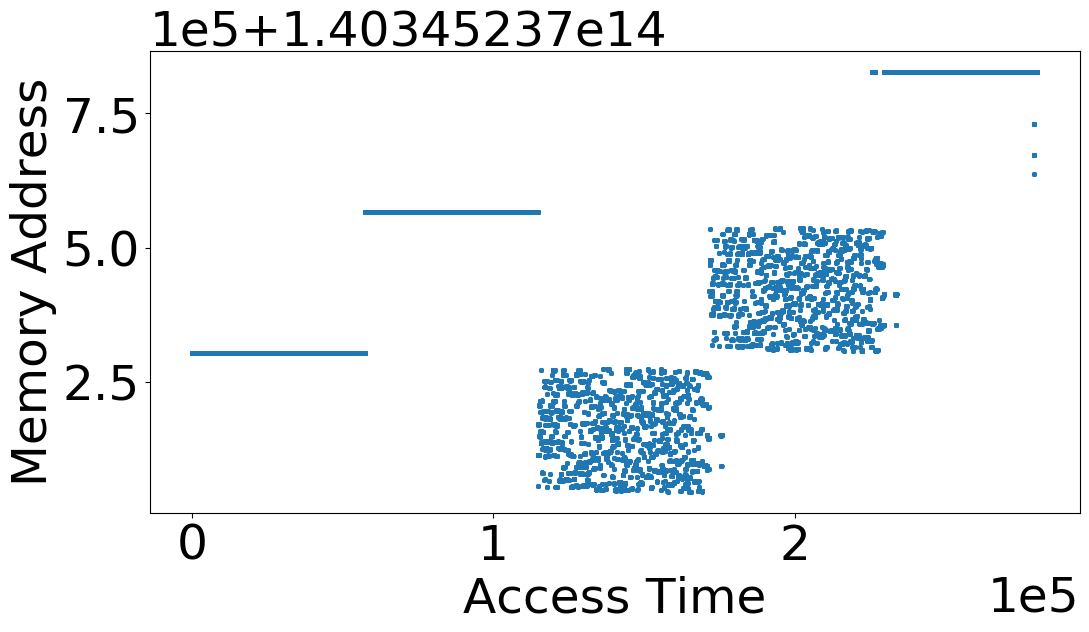}
	}
	\subfloat[SVM (I)]{
		\includegraphics[width=0.24\textwidth]{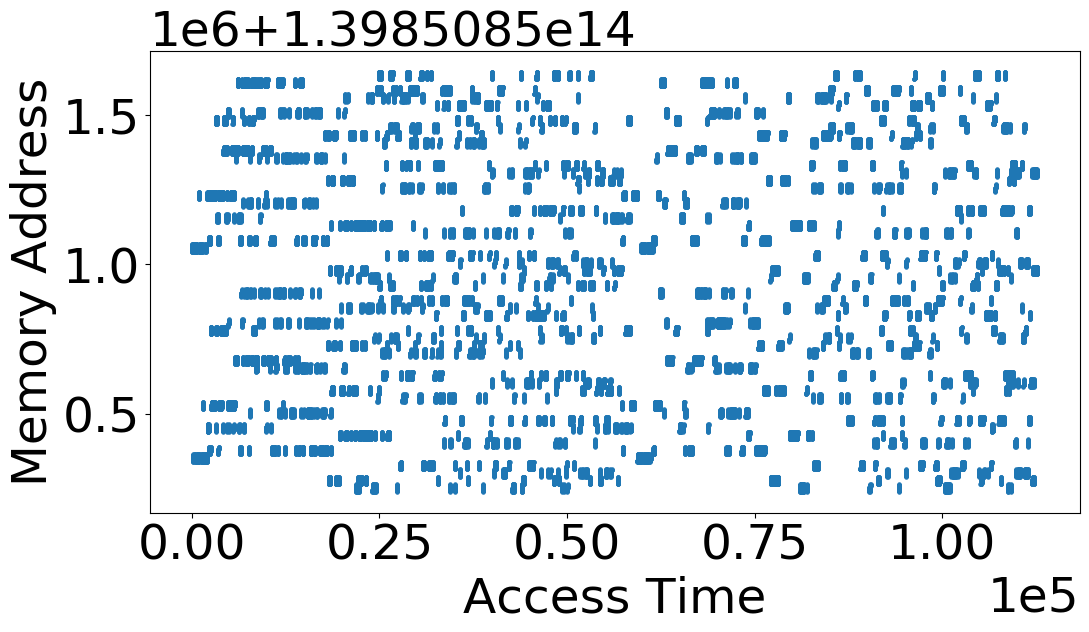}
	}
	\subfloat[SYR2K (I)]{
		\includegraphics[width=0.24\textwidth]{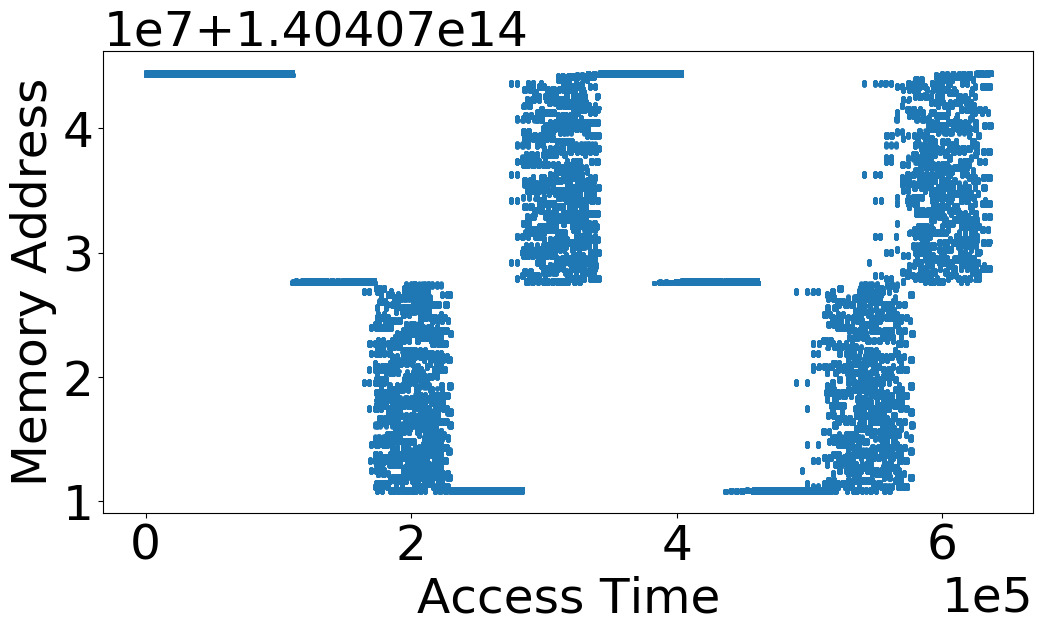}
	}
	\subfloat[SYR2K (R)]{
		\includegraphics[width=0.24\textwidth]{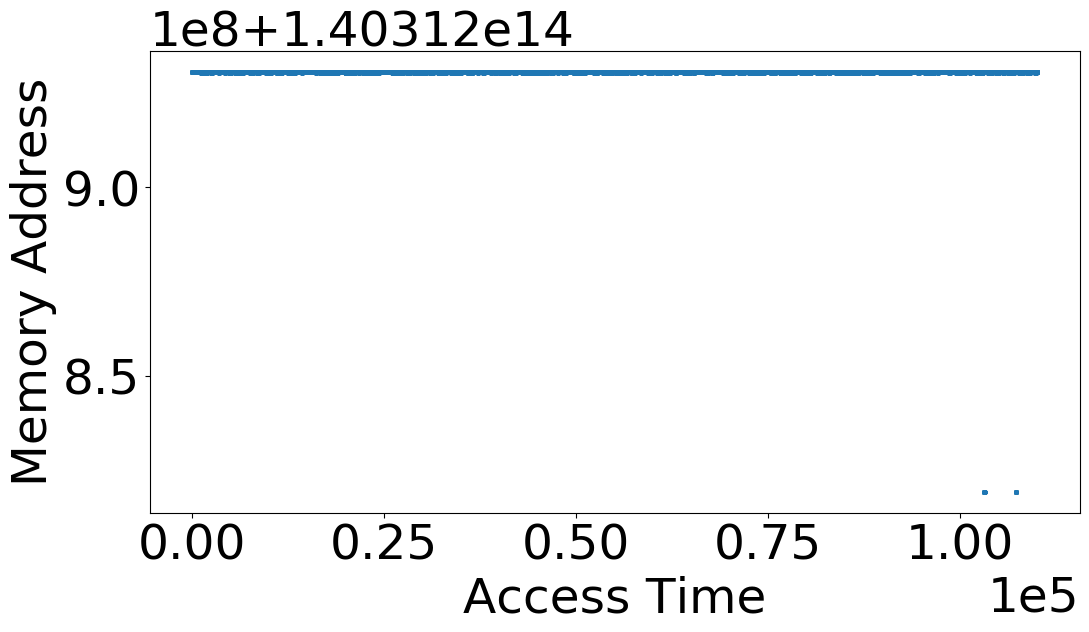}
	}
	\caption{Memory access patterns of benchmarks in UVMBench (continued).}
	\label{access_pattern}
\end{figure*}

\subsection{Memory Access Pattern Profiling}
\label{pattern}
To study the relationship between memory behaviors and UVM efficiency, we first profile memory access patterns of each benchmark. In this experiment, NVBit is used to generate memory reference traces by injecting the instrumentation function before performing each global load/store. The memory traces are plotted in Figure \ref{access_pattern}. The horizontal axis corresponds to the logical access time, and the vertical axis shows the accessed memory addresses.

\begin{figure*}[t]
	\centering
	\includegraphics[width=\linewidth]{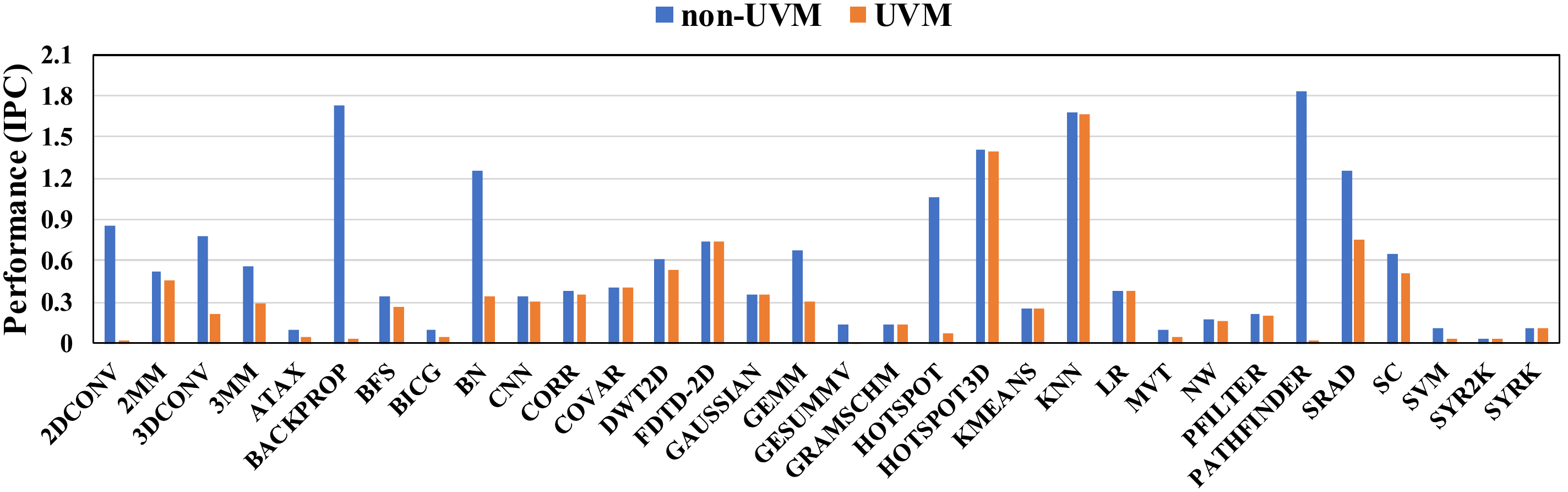}
	\caption{Direct UVM conversion in \projname \ leads to large performance degradation vs. non-UVM. }
	\label{ipc}
\end{figure*}

As can be seen from the figure, benchmarks in the \projname \ suite exhibit diverse memory access patterns. They can be generally classified into \textit{regular} and \textit{irregular} memory access patterns, as indicated after each benchmark name as (R) or (I) in Figure \ref{access_pattern} (and as indicated in the ``Type'' column in Table \ref{benchname}). This classification follows the same classification method as \cite{li2019framework}: if benchmarks access only a small number of memory pages at any point of time, they are classified as \textit{regular} benchmarks; in contrast, benchmarks with large unique memory pages access at a given time are identified as \textit{irregular} benchmarks. For regular benchmarks (e.g., 2DCONV, 2MM and so on), they exhibit a streaming access pattern. These benchmarks access only a small number of memory addresses and seldom exhibit data reuse within the kernel. In contrast, irregular benchmarks show very different memory access patterns: accessing many memory addresses at a given time (e.g., ATAX, BICG, GAUSSIAN), repeatedly accessing the same memory address over time (e.g., COVAR, GRAMSCHM), or accessing random addresses (e.g., SC, SVM). Note that benchmark \textit{NW} is classified as irregular, as it exhibits a sparse, localized and repeated memory accesses, although this is not quite visible in the figure due to the scale. In the experiment of memory oversubscription presented later in Section \ref{oversub_part}, we find that benchmark performance is highly related to memory access patterns.       

\subsection{UVM \textit{vs.} non-UVM Performance}

\noindent\textit{a. Performance of Direct UVM Conversion}

As mentioned earlier, while UVM greatly eases programming efforts by removing explicit memory management, this is achieved at the cost of certain performance overhead, particularly with naive/direct conversion to UVM. Figure \ref{ipc} compares the performance of all the benchmarks in the non-UVM and UVM programming models. The IPCs are obtained from Nvidia \textit{nvprof}. Across the benchmarks, the performance of the UVM version has an average of 34.2\% slowdown compared with the non-UVM one. These results are expected as the page fault handling causes large performance overhead for kernel execution. Under the UVM programming model, data is allowed to reside in other location (e.g., on the CPU side) while a kernel is executing. When the required data does not reside in the GPU DRAM (page fault occurrence), the kernel has to be stalled while waiting for the data to be fetched from the CPU side. In the non-UVM version, programmers have made sure that data is always available on the GPU side.


Among these benchmarks, we can observe that \textit{2DCONV}, \textit{BACKPROP}, \textit{HOTSPOT}, \textit{GESUMMV} and \textit{PATHFINDER} have the most significant performance drop in the UVM implementation. The reason is that, for these 5 benchmarks, the data migration time accounts for majority of the entire execution (over 80\%), and their kernels have little to no data reuse and are only invoked once. A considerable amount of stall time occurs during the one-time execution of the kernels to wait for data, and the fetched data is not used again.
These factors lead to the observed large performance degradation. However, as shown shortly, the performance degradation can be greatly mitigated with some additional programming efforts. 

\begin{figure}[t]
	\centering
	\includegraphics[width=\linewidth]{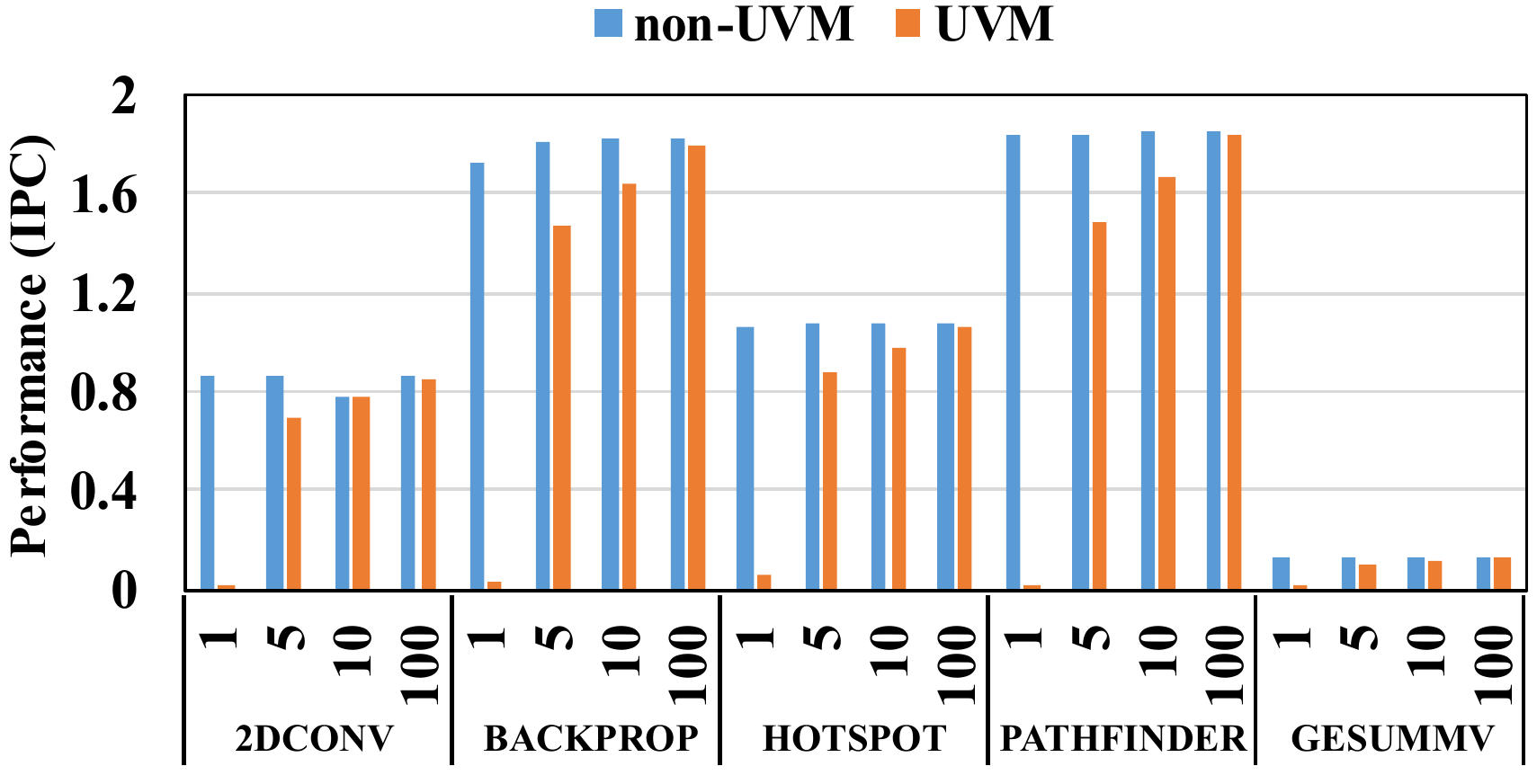}
	\caption{Performance of UVM restores with increased number of kernel invocations.}
	\label{ipc_in}
\end{figure}

\vspace{1ex}
\noindent\textit{b. Restoring UVM Performance via Data Reuse}

Data reuse can mitigate UVM performance degradation by reducing the occurrence of page faults. As mentioned earlier, we study the impact of data reuse by modifying the number of times a kernel is invoked. Figure \ref{ipc_in} plots the change in performance as we increase the kernel invocation times (there is no kernel execution dependency between consecutively invoked kernels). It can be seen that the performance of these benchmarks under UVM is rapidly improving with more invocation and eventually approaches to the performance of non-UVM. Except for the first executed kernel, the following kernels in the GPU program may reuse the data that has been fetched during the execution of the first kernel, and fewer page faults would occur. The results confirm that more data reuse leads to smaller data migration overhead. 

\noindent \textbf{\textit{Observation/Suggestion:}} Although data reuse is artificially introduced in the software program in this experiment, it prompts us that if applications exhibit significant data reuse opportunities, either inherent or created through architecture optimizations, UVM can be an attractive model that provides flexibility while having little performance overhead.



\vspace{1ex}
\noindent\textit{c. Restoring UVM Performance via Data Prefetch}

\begin{figure}[t]
	\centering
	\includegraphics[width=\linewidth]{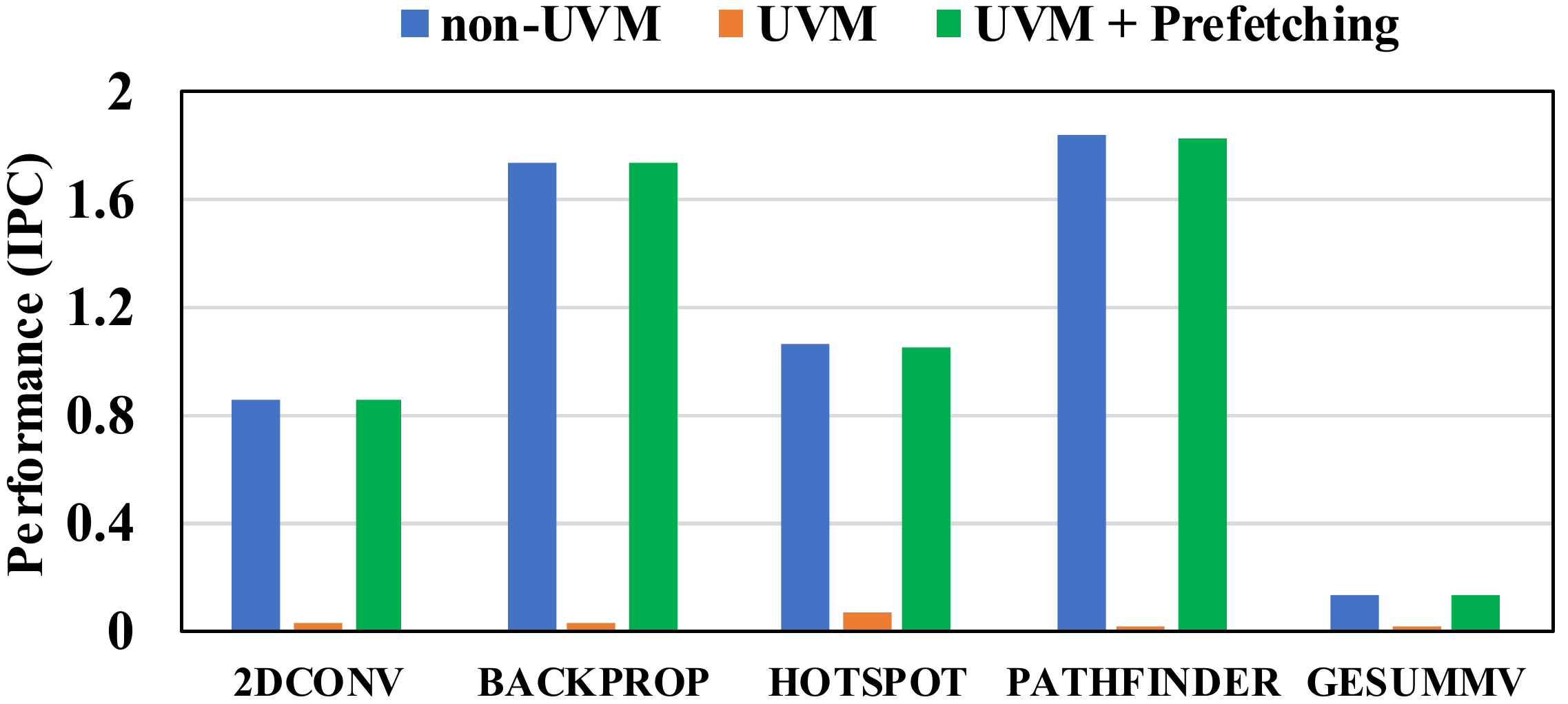}
	\caption{Performance of UVM restores by enabling prefetching.}
	\label{ipc_pre}
\end{figure}

Nvidia provides a runtime API \textit{cudaMemPrefetchAsync} that enables asynchronous data prefetching. Through this API, data can be prefetched to the device memory before the data is accessed by a kernel on the GPU. This reduces the occurrence of page faults. To study the impact of prefetching on UVM kernel execution performance, we augment all the benchmarks in \projname \ with such prefetching capability. 
Figure \ref{ipc_pre} shows the results from the above 5 benchmarks that experience the largest performance drop in UVM.

It can be observed that the performance of these benchmarks improves considerably after this optimization and is close to the performance of the non-UVM version. The geometric mean of the slowdown has decreased from 95.8\% to merely 0.7\%. The improvement comes from the fact that kernel execution is now rarely stalled as data has already been fetched in the device memory before being accessed. 
While not shown, the performance of other 27 UVM-version of the benchmarks also restores to very close to the non-UVM version after using asynchronous prefetching. 


\noindent \textbf{\textit{Observation/Suggestion:}} 
Besides data resue, another alternative to restore performance degradation of UVm is data prefetching by employing the runtime API \textit{cudaAsyncPrefetch}. In theory, page faults can be completely eliminated if there is an oracle prefetecher that is able to load any required data into the GPU memory before the data is accessed. That can serve as an upper-bound of future UVM prefetech schemes.


\ul{It is important to note that, we achieve data reuse and data prefetch in the above experiments by manually modifying the software programs.} In other words, these optimizations are realized on the software side and requires additional programming efforts. This is not the intention of UVM that aims to reduce programming efforts. 
In practice, what is needed is innovation in architecture research that can achieve similar level of data reuse and prefetch but is transparent to programmers. Facilitating research along this line is what our \projname \ suite is created for.


\begin{figure*}[t]
	\centering
	\includegraphics[width=\linewidth]{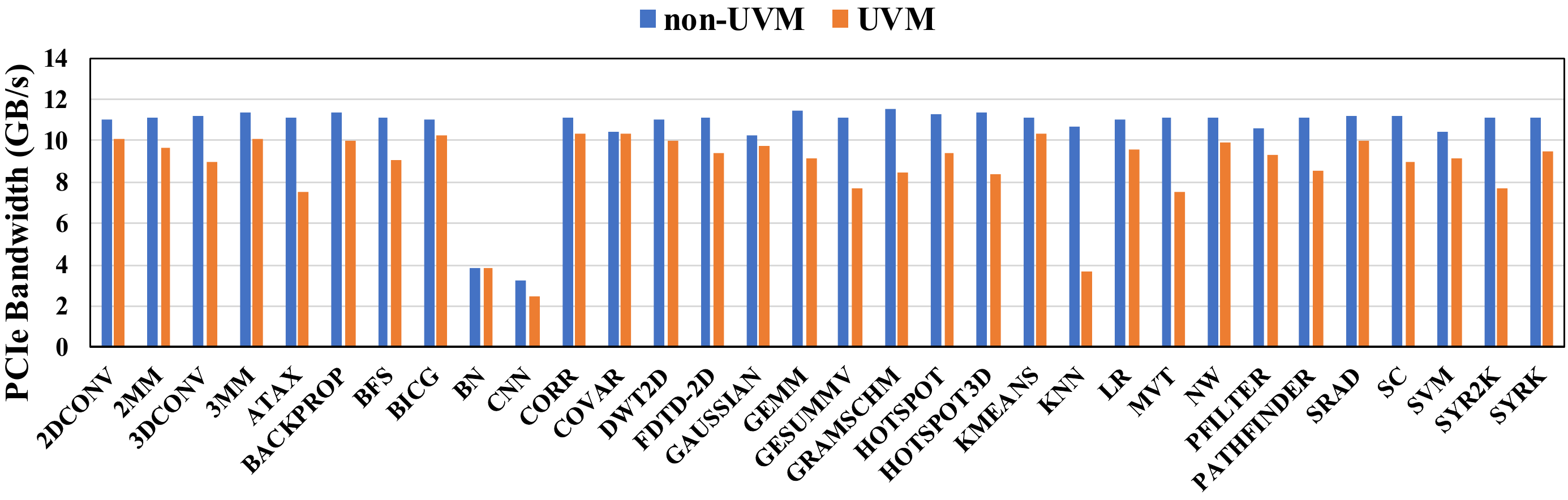}
	\caption{Achieved PCIe bandwidth of non-UVM vs. UVM during data migration.}
	\label{pcie}
\end{figure*}

\begin{figure*}[t]
	\centering
	\includegraphics[width=\linewidth]{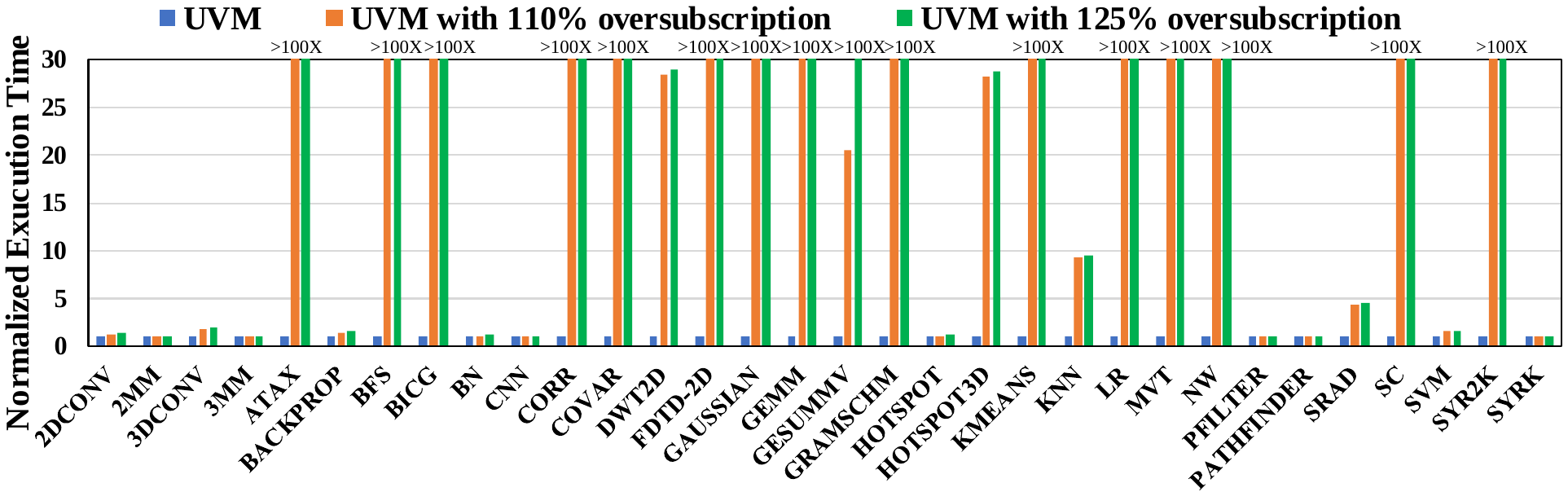}
	\caption{Change in benchmark execution time when GPU memory oversubscripted (normalized to no memory oversubscription).}
	\label{oversub}
\end{figure*}

\subsection{Effect of Data Migration on PCIe Bandwidth}

The performance of data migration between CPU and GPU also closely relates to the effective PCIe bandwidth. Under the UMV programming model, variable sized on-demand data is transferred from the CPU memory to the device memory. To understand performance trade-offs, it is worth studying the effect of UVM data migration on the PCIe link. 
Figure \ref{pcie} compares the achieved PCIe bandwidth with non-UVM and UVM programming models during data migration. On average, the achieved PCIe bandwidth of UVM is 15.2\% lower than that of non-UVM. In general, the larger the transferred data size is, the higher the effective PCIe bandwidth can achieve. This is mainly because of the constant PCIe protocol overhead and limited hardware resources (e.g., data buffer size, number of DMA channels, number of outstanding requests, etc.), so the overhead can be amortized better with larger transferred data. Since the non-UVM model copies the entire allocated data chunk to the GPU memory before execution, this results in relatively high effective bandwidth. In contrast, the migrated data size in UVM is usually much smaller than the non-UVM one as only on-demand data is migrated through the PCIe bus (usually smaller than 1MB). Note that benchmarks \textit{BN} and \textit{CNN} in UVM and non-UVM both exhibit low effective PCIe bandwidth, because the sizes of allocated variables in these two benchmarks are all small (less than 4KB), and even the entire chunk of allocated variable transmission cannot fully utilize the PCIe bandwidth. 



Figure \ref{pcie} also shows that, among UVM benchmarks, the effective PCIe bandwidth may vary a lot. The variation is mainly caused by the hardware prefetcher inside the GPU. For example, Nvidia has implemented a tree-based hardware prefetcher in their GPUs, which heuristically adjusts the prefetching granularity based on access locality. The difference in memory access patterns across benchmarks put the hardware prefetcher in different degrees of efficacy. More detailed discussion on UVM hardware prefetchers can be found in other papers such as \cite{ganguly2019interplay, yu2019quantitative, li2019framework}.

\noindent \textbf{\textit{Observation/Suggestion:}} The above results on the effective PCIe bandwidth indicate that hardware prefetchers that are currently employed in GPUs cannot fully utilize PCIe bandwidth. Thus, future research is much needed to continue developing and optimizing GPU hardware prefetchers that are UVM-aware.


\subsection{Oversubscription}
\label{oversub_part}


A major advantage of UVM is to enable kernel execution when memory is oversubscribed. Performance under memory oversubscription can be significantly reduced since part of the data now needs to be brought from the CPU memory. Despite this, UVM is still very attractive, as such memory oversubscription is not possible under non-UVM.
To quantify the performance degradation when the GPU memory is oversubscribed, we run all the benchmarks in the suite under various memory capacities. As different benchmarks have different required memory footprint, to create memory oversubscription, we modify the available memory space through the \textit{cudaMalloc} runtime API. The required memory footprint is set to be 110\% and 125\% of the available memory space in the GPU physical memory.
Figure \ref{oversub} shows the results. 
As expected, all the benchmarks suffer considerable performance degradation under memory oversubscription. The more memory is oversubscribed, the more performance degrades. 

From Figure \ref{oversub}, we also observe that many of the benchmarks can complete execution with 2-3x slowdown under memory oversubscription, whereas other benchmarks suffer from a significant performance penalty or even crash, marked as $>$100X in the figure (e.g., LR uses the cublas library which cannot support memory oversubscription and leads to crash).
For the former, we find that the main performance overhead is caused by kernel stalls when waiting for the eviction of pages to create space for newly fetched data. These benchmarks usually have a streaming access pattern (Section \ref{pattern}). With this pattern and the LRU eviction policy in Nvidia GPUs, the evicted data does not affect kernel execution as the evicted data is not reused any more. Therefore, the performance overhead mainly comes from the waiting time of page eviction. 
For the latter, 
the large performance penalty mainly comes from severe page thrashings, which repeatedly migrate the page back and forth between the GPU and the CPU. This usually occurs when a benchmark has a short data reuse distance so the evicted data is needed/reused within a short time. 
Note that, although the degradation seems large, 
UVM is still much better non-UVM which does not allow kernels to run at all if the memory is oversubscribed.

\noindent \textbf{\textit{Observation/Suggestion:}} The significant performance degradation under memory oversubscription suggests that the current eviction policies are doing a poor job at selecting the best candidate pages to evict, thus causing severe page thrashings and limiting the amount of memory that can be oversubscribed. This may be possibly because existing eviction policies are not designed specifically with supporting UVM in mind. We urge researchers to develop more effective eviction policies that can select evicted data more accurately or even proactively to make space for expected data accesses.

\section{Conclusion}
\label{conclusion}
The Unified Virtual Memory (UVM) programming model has been introduced recently in GPUs to ease the programming efforts and to allow kernel execution under memory oversubscription. This paper identifies the need for representative benchmarks for GPU UVM, and proposes a comprehensive benchmark suite to help researchers understand and study various aspects of GPU UVM. Several observations and suggestions have been drawn from the evaluation results to guide the much needed future research on UVM.




\end{document}